\documentclass[Conference,a4paper]{IEEEtran}
\IEEEoverridecommandlockouts
\usepackage{color}
\usepackage{graphicx}
\usepackage{epstopdf}
\usepackage{amsmath}
\usepackage{amssymb}
\usepackage{algorithm}
\usepackage{algorithmic}
\usepackage{amsmath}
\usepackage{multirow}
\usepackage{booktabs}
\usepackage{array}
\usepackage{amsthm}
\usepackage{stfloats}
\usepackage{caption}
\usepackage{subfigure}
\usepackage{bm}
\usepackage{booktabs}
\usepackage{setspace}
\usepackage{diagbox}
\usepackage{enumerate}
\usepackage{ulem}
\usepackage{url}
\usepackage{circledsteps}

\usepackage{hyperref}
\allowdisplaybreaks[4]

{\bgroup
 \addtolength\abovedisplayshortskip{#1}
 \addtolength\abovedisplayskip{#1}
 \addtolength\belowdisplayshortskip{#1}
 \addtolength\belowdisplayskip{#1}
 }
{\egroup\ignorespacesafterend}

\newtheorem{lemma}{Lemma}

\newcommand{\be}{\begin{equation}}
\newcommand{\ee}{\end{equation}}
\newcommand{\bea}{\begin{eqnarray}}
\newcommand{\eea}{\end{eqnarray}}
\newcommand{\ba}{\begin{array}}
\newcommand{\ea}{\end{array}}

\title{
Transmit Beamforming Design for Integrated Sensing and Communication Using Transmissive RIS Transceiver
}
\author{\IEEEauthorblockN{Yuan Guo, Wen Chen, Yang Liu, Qiong Wu, and Weiren Zhu
}
\thanks{
Y. Guo, W. Chen,
and W. Zhu
are with the Department of Electronic Engineering, Shanghai Jiao Tong University, Shanghai 200240, China 
(e-mail:
yuanguo26@sjtu.edu.cn;
wenchen@sjtu.edu.cn;
weiren.zhu@sjtu.edu.cn).}
\thanks{
Y. Liu is with the School of Information and Communication Engineering, Dalian University of Technology, Dalian 116024, China
(e-mail: 
yangliu\_613@dlut.edu.cn).}
\thanks{
Q. Wu is with the School of Internet of Things Engineering, Jiangnan University, Wuxi 214122, China
(e-mail: 
qiongwu@jiangnan.edu.cn).
}
}

\begin{document}
\maketitle
\pagestyle{empty}
\thispagestyle{empty}

\begin{abstract}
Integrated sensing and communication (ISAC) is a key technology for future wireless networks, 
calling for hardware-efficient architectures to jointly support communication and sensing. 
In this paper, 
a transmissive reconfigurable intelligent surface (TRIS) transceiver is leveraged to enable an ISAC system.
Under the considered system model,
we investigate transmit beamforming design for the TRIS transceiver 
to maximize the sum-rate/beampattern gain, 
subject to the predefined sensing beampattern gain/communication rate thresholds 
and the per-unit power constraints of the TRIS transceiver.
Since the objective functions and constraints are non-convex,
the above two optimization problems are highly challenging.
To resolve the difficult optimization problems,
we combine the fractional programming (FP) method and the majorization-minimization (MM) framework to develop second-order cone programming (SOCP)-based solutions.
Since the per-element power constraints introduce a large number of constraints, 
this increases the complexity of solving the optimization problems.
By splitting the coupling constraints and applying the alternating direction method of multipliers (ADMM) framework,
we propose two analytic-based algorithms for efficiently updating the beamformer configurations in the sum-rate and beampattern gain maximization problems, respectively. 
Simulation results demonstrate the convergence and effectiveness of the proposed algorithms,
and show that the low-complexity algorithms achieve performance close to the SOCP-based benchmarks with substantially reduced computational complexity.

\end{abstract}


\maketitle
\section{Introduction}

{
The upcoming sixth-generation (6G) wireless communication networks are expected to enable diverse services and applications, 
including low-altitude communication networks and vehicle-to-everything (V2X) communications, 
with significantly improved communication and sensing capabilities \cite{ref_6G_1}.}
As a cornerstone technology for the 6G wireless network,
integrated sensing and communication (ISAC) has attracted considerable attention from both academia and industry worldwide \cite{ref_ISAC_1}$-$\cite{ref_ISAC_2}.
Specifically, 
the ISAC technology can jointly realize sensing and communication functionalities on a unified hardware platform,
effectively reducing hardware costs and improving spectral efficiency.
Furthermore, 
communication and sensing can be jointly designed to realize mutual gains instead of being treated as separate objectives.
In this context, ISAC has been envisioned as a promising solution by effectively integrating dual capabilities and leveraging their mutual assistance.
{
Thus,
many recent advances in joint radar sensing and communication design can be found in \cite{ref_ISAC_1}$-$\cite{ref_ISAC_5} and the references therein.}

Recently, the ability to reconfigure wireless channels has emerged as a pivotal research direction in wireless network design.
The emerging reconfigurable intelligent surface (RIS) \cite{ref_RIS_1}$-$\cite{ref_RIS_2} is envisioned as a representative technology.
Generally,
RIS is a planar surface composed of a multitude of low-cost and passive tunable units,
each of which can dynamically adjust the phase shifts of the electromagnetic (EM) waveforms 
to enhance reflected signal strength and provide substantial beamforming gain.
Additionally, 
RIS can be installed near the transmitter, near the receiver, or in the environment to realize performance improvements.
This attractive feature has driven rapid growth in research into the diverse applications of RIS technology, as reported in \cite{ref_RIS_1}$-$\cite{ref_RIS_5} and the references therein.
{ Specifically,
recent works on the application of RIS in ISAC systems can be found in \cite{ref_RIS_ISAC_2}$-$\cite{ref_RIS_ISAC_4}, 
together with the references therein.

Note that the aforementioned works \cite{ref_RIS_1}$-$\cite{ref_RIS_ISAC_4} focus on various RIS architectures,
i.e., conventional RIS \cite{ref_RIS_1}$-$\cite{ref_RIS_ISAC_3},
active RIS \cite{ref_RIS_ISAC_3_1},
and 
simultaneously transmitting and reflecting (STAR)-RIS \cite{ref_RIS_ISAC_4},
which serves as an auxiliary in wireless networks.}
{ Very recently,
a novel transceiver architecture named \textit{transmissive RIS (TRIS) transceiver} has been introduced in \cite{ref_TRIS_1}.
This TRIS transceiver architecture mainly consists of a passive TRIS and a single horn antenna feed,
which can avoid using numerous radio frequency (RF) chains and complex signal processing units,
while achieving comparable system performance with lower power consumption and lower cost.}
In addition, 
compared with the reflective RIS transmitters detailed in \cite{ref_RIS_transmitter_1}, 
the TRIS transceiver can mitigate feed-source blockage and echo-interference issues.

Therefore,
the potential of the TRIS transceiver has been extensively explored in many emerging applications \cite{ref_TRIS_app_1}$-$\cite{ref_TRIS_app_10}.
For instance,
the authors of \cite{ref_TRIS_app_1} investigated a multi-stream downlink communication scheme using the TRIS transceiver,
which leverages time-modulated array (TMA) technology to achieve higher-order modulation and multi-stream beamforming.
Besides, a linear-complexity beamforming solution was proposed for solving the non-convex signal-to-interference-plus-noise ratio (SINR) fairness problem.
The paper \cite{ref_TRIS_app_2} considered a TRIS transceiver-enabled simultaneous wireless information and power transfer (SWIPT) system and investigated a sum-rate maximization problem under imperfect channel state information (CSI).
The work \cite{ref_TRIS_app_3} adopted the TRIS transceiver to improve computing capability, reduce computing delay, and lower transmitter deployment cost in a multi-tier computing network.
The literature \cite{ref_TRIS_app_4} designed a novel hybrid active-passive RIS transmitter architecture in which each unit can flexibly switch between active and passive modes. 
Based on this architecture, 
the authors formulated and solved an energy efficiency (EE) maximization problem and validated the superiority of the proposed design.
The paper \cite{ref_TRIS_app_5} studied robust beamforming for a secrecy multi-user multiple-input single-output (MISO) network assisted by both a TRIS transceiver and an RIS, accounting for imperfect reflection CSI.
Taking into account imperfect CSI and modeling the channels of unauthorized users,
the authors in \cite{ref_TRIS_app_6} investigated a robust and secure ISAC system enhanced by a TRIS transceiver and rate-splitting multiple access (RSMA) technology.
{In \cite{ref_TRIS_app_7}, 
a distributed cooperative ISAC system aided by a TRIS transceiver was developed for enlarging coverage 
and improving wireless environment awareness.}
The authors of \cite{ref_TRIS_app_8} leveraged a TRIS transceiver together with the signal-level spatial registration algorithm to enable a cooperative integrated sensing, computing, and communication (ISCC) network 
that achieves diverse functionalities with low energy consumption.
For a TRIS transceiver-enabled multi-group multi-cast downlink communication system,
the paper \cite{ref_TRIS_app_9} designed three optimization algorithms to address the considered rate fairness problem.
The work \cite{ref_TRIS_app_10} investigated a TRIS transceiver-empowered SWIPT system.

\subsection{Motivations and Contributions}

{
While prior works \cite{ref_TRIS_app_1}$-$\cite{ref_TRIS_app_10} have studied various applications of TRIS transceivers, 
their integration into ISAC systems has received only limited attention.
Moreover, conventional ISAC transmitters are typically built on multi-antenna architectures that require multiple RF chains and sophisticated signal processing modules to support both communication and sensing functions, resulting in high hardware cost, power consumption, and implementation complexity. 
In contrast, the TRIS transceiver can substantially reduce RF-chain usage and signal processing complexity while still enabling flexible transmit beamforming, 
making it an appealing low-power and low-cost architecture for ISAC.
On the other hand, 
traditional sensing metrics, 
such as the Cram\'er-Rao bound (CRB), mean squared error (MSE), and detection probability (DP), 
often suffer from mathematical intractability or rely on specific estimators or detectors.
By contrast, beampattern gain is physically interpretable and mathematically tractable, 
and directly reflects the transmit-side illumination power toward the target. 
Therefore,
we adopt the target beampattern gain as the sensing metric.
Although papers \cite{ref_TRIS_app_6}$-$\cite{ref_TRIS_app_7} investigated TRIS transceiver-enabled ISAC systems, 
beampattern-oriented sensing design for such systems remains largely unexplored.
Motivated by these observations,
in this paper, 
we study a TRIS transceiver-enabled ISAC system using target beampattern gain as the sensing metric.}
Specifically,
the key contributions of this paper are summarized as follows:
\begin{itemize}
\item
This paper studies the beamforming design in an ISAC system enhanced by a TRIS transceiver to achieve simultaneous communication and target sensing.
The objective of this paper is to maximize the sum-rate/beampattern gain by optimizing the communication and radar beamformers,
subject to the sensing beampattern gain/individual communication rate thresholds and the per-element power required by the TRIS transceiver.
Note that the sensing beampattern gain criterion is rarely investigated in the ISAC system with the TRIS transceiver.

\item
Due to the non-convexity of both the objective and the constraints, 
the considered problems (e.g., sum-rate maximization and beampattern gain maximization) are highly challenging.
Specifically,
in the sum-rate maximization problem, 
the achievable rate objective is non-concave, 
and the sensing requirement is non-convex. 
In the beampattern gain maximization problem, 
the quadratic beampattern gain is maximized directly, 
while the individual rate constraints remain non-convex. 
To facilitate efficient solutions for the aforementioned problems,
the rate function is first reformulated into an equivalent form using the fractional programming (FP) framework \cite{ref_FP}$-$\cite{ref_FP_1}.
Subsequently,
by employing the majorization-minimization (MM) technique \cite{ref_MM}, 
the original non-convex optimization problems can be transformed into second-order cone programming (SOCP) \cite{ref_Convex Optimization} formulations that can be efficiently solved at each iteration.

\item
Although the per-element power constraints are convex, 
their large number significantly increases the computational complexity of repeatedly solving the resulting SOCP subproblems within the iterative optimization procedure.
Therefore,
by splitting the coupling constraints and leveraging the alternating direction method of multipliers (ADMM) method \cite{ref_ADMM},
we successfully develop two closed-form solutions that can effectively update the beamformer configurations of the sum-rate and beampattern gain maximization problems, respectively.

\item
Last but not least,
extensive numerical results are presented to verify the effectiveness and efficiency of our proposed solutions under various system configurations.
Besides,
the results also show that the low-complexity algorithms achieve lower computational complexity than the SOCP-based approaches while maintaining the system performance.

\end{itemize}

\section{System Model and Problem Formulation}
\subsection{System Model}

\begin{figure}[t]
	\centering
	\includegraphics[width=.35\textwidth]{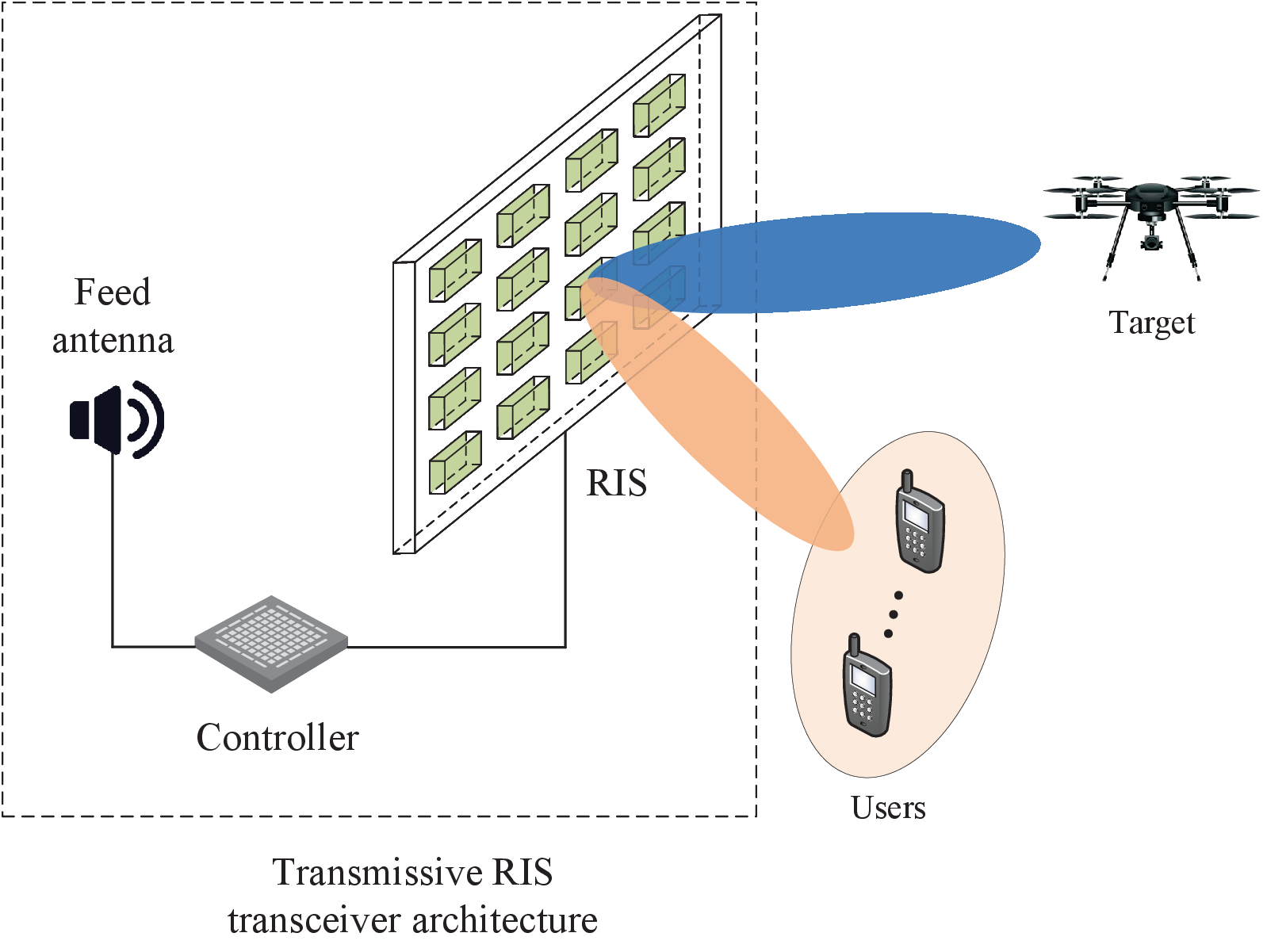}
	\caption{An illustration of the TRIS transceiver-enabled ISAC system.}
	\label{fig.1}
\end{figure}

As shown in Fig. \ref{fig.1},
we investigate a TRIS transceiver-enabled ISAC system 
that comprises a TRIS transceiver equipped with $N$ units, 
$K$ single-antenna mobile users,
and one point-like target.\footnote{The proposed algorithms in this manuscript can also be applied to multi-target and multi-clutter 
 scenarios with slight modifications.}
For convenience, 
the sets of mobile users and TRIS transceiver units are denoted as 
$\mathcal{K}$
and 
$\mathcal{N}$,
respectively.
The considered system can be interpreted as a compact 6G ISAC access node, 
where a TRIS transceiver simultaneously serves multiple mobile users and senses a target of interest in the surrounding area. 
Representative application scenarios include low-altitude wireless access, 
roadside units in V2X networks, 
and local hotspot transceivers, 
where both communication service and environmental awareness are required. 
Therefore, 
the point-like target adopted in our model serves as a tractable abstraction for an object to be sensed, 
such as a UAV, vehicle, or other mobile object.

First, 
by adopting the TMA technique \cite{ref_TMA},\footnote{ 
{ As the core implementation technique of the TRIS transceiver, 
TMA enables a single RF feed to realize multiple equivalent harmonic-domain beamforming coefficients 
through periodic time modulation. 
Specifically, 
the horn antenna illuminates the TRIS elements with a single carrier signal, 
and each TRIS element applies a controllable periodic transmission coefficient to the incident signal. 
Due to the periodic time variation, 
the transmitted signal contains multiple frequency-shifted harmonic components around the carrier frequency, 
whose amplitudes and phases are determined by the Fourier coefficients of the corresponding periodic transmission coefficients. 
Therefore, 
the beamforming vectors can be interpreted as harmonic-domain coefficients generated by the time-modulated TRIS elements.}} 
the TRIS transceiver generates the communication and probing waveforms 
and multiplexes them via superposition multiplexing.
{
Accordingly,
the general transmit signal 
at the TRIS transceiver is expressed as}
\begin{align}
\mathbf{x} = {\sum}_{k=1}^{K} \mathbf{w}_{c,k}x_{c,k}
+
{\sum}_{n=1}^{N} \mathbf{w}_{r,n}x_{r,n}, \label{Equivalent baseband model}
\end{align}
where
$x_{c,k} \in \mathbb{C}$ is the modulated information signal 
that is assumed to be independent and identically distributed (i.i.d.) 
circularly symmetric complex Gaussian (CSCG) random variable with zero mean and unit variance,
i.e., $x_{c,k} \sim  \mathcal{CN}(0,1)$,
and $\mathbf{w}_{c,k}\in \mathbb{C}^{N\times 1}$ represents its corresponding communication beamforming.
Similarly,
$x_{r,n} \in \mathbb{C}$ is the modulated radar signal
satisfying
$\mathbb{E}\{  x_{r,n}   \} = 0$
and
$\mathbb{E}\{ \vert x_{r,n}  \vert^2 \} = 1$,
and the vector 
$\mathbf{w}_{r,n}\in \mathbb{C}^{N\times 1}$
denotes the corresponding radar beamforming.
Besides,
we assume that communication and radar signals are statistically independent and uncorrelated, 
e.g., $\mathbb{E}\{   x_{c,k}x_{r,n}^{\ast}   \} = 0, \forall k \in \mathcal{K}, \forall n \in \mathcal{N}$.

As proposed in \cite{ref_TRIS_1},
the TRIS transceiver consists of a feed horn antenna, 
a control unit, 
and a TRIS made from reconfigurable elements arranged in a uniform planar array (UPA).
This novel architecture provides a multi-stream communication solution 
that is energy-efficient 
and well-suited to rapidly changing propagation conditions and deployments with limited power availability.
For notational compactness and unified algorithmic derivation, 
we define 
$\mathbf{w}_{c} \triangleq [\mathbf{w}_{c,1}^T, \cdots, \mathbf{w}_{c,K}^T]^T \in \mathbb{C}^{NK\times 1}$
and
$\mathbf{w}_{r} \triangleq [\mathbf{w}_{r,1}^T, \cdots, \mathbf{w}_{r,N}^T]^T \in \mathbb{C}^{NN\times 1}$.
Based on the TMA technique detailed in \cite{ref_TRIS_app_1},  
the beamforming vectors are required to satisfy per-element power constraints,
which can be written as
\begin{align}
\mathbf{w}_{c}^H \bar{\mathbf{A}}_{c,n} \mathbf{w}_{c} 
+ \mathbf{w}_{r}^H \bar{\mathbf{A}}_{r,n}\mathbf{w}_{r} \leq P_t, \forall n \in \mathcal{N},
\end{align}
where 
$P_t$ denotes the maximum power limit for each TRIS transceiver unit,
and the newly introduced selection matrices 
$\bar{\mathbf{A}}_{c,n}$ and  $\bar{\mathbf{A}}_{r,n}$
are respectively expressed as
\begin{align}
&\bar{\mathbf{A}}_{c,n} \triangleq \text{blkdiag}(\underbrace{\mathbf{A}_n, \cdots,\mathbf{A}_n}
\limits_{K}
  ) \in \mathbb{R}^{NK\times NK},\\
&\bar{\mathbf{A}}_{r,n} \triangleq \text{blkdiag}(\underbrace{\mathbf{A}_n, \cdots,\mathbf{A}_n}
\limits_{N}
  ) \in \mathbb{R}^{NN\times NN},
\end{align}
and the above matrix $\mathbf{A}_n$ can be defined as
$\mathbf{A}_n \triangleq \text{diag}(\mathbf{a}_{n}) \in \mathbb{R}^{N\times N}$,
and the index vector $\mathbf{a}_{n}$ is formulated as
\begin{align}
\mathbf{a}_{n} \triangleq [0,0,\cdots,\underbrace{1}\limits_{\textrm{n-}th},\cdots,0,0]^T \in \mathbb{R}^{N\times 1},
\end{align}
where the element at position 
$n$ equals $1$ and all other elements are $0$.

\textit{1) Communication Model:}
In this work, we assume all channels follow a quasi-static block-fading model 
and focus on a specific fading block during which the relevant channels are assumed to be constant.
Let $\mathbf{h}_{c,k} \in \mathbb{C}^{N\times 1}$ represent the equivalent baseband channel from the TRIS transceiver to the $k$-th user.
To investigate the performance upper bound of the considered TRIS transceiver-empowered ISAC system, 
we consider that the TRIS transceiver has perfect knowledge of CSI for all relevant links,\footnote{{
Although imperfect CSI and target-location uncertainty are more practical, 
this paper focuses on the performance upper bound of the considered TRIS transceiver-enabled ISAC system. 
A similar assumption has also been adopted in \cite{ref_TRIS_app_1}, \cite{ref_TRIS_app_3}, and  \cite{ref_TRIS_app_7}.
Besides,
several methods for CSI acquisition in TRIS transceiver-enabled systems have been developed in \cite{ref_channel_estimation_1}$-$\cite{ref_channel_estimation_2}.
We leave the case of imperfect CSI and target-location uncertainty as an important topic for future work.
}
}
following the channel estimation methods in \cite{ref_channel_estimation_1}$-$\cite{ref_channel_estimation_2}.
The received signal at the $k$-th user is given by
\begin{align}
{y}_{c,k}
&=
\underbrace{{\mathbf{h}}_{c,k}^H\mathbf{w}_{c,k}x_{c,k}}
\limits_{\textrm{Desired signal}}
+\underbrace{{\sum}_{i\neq k}^{K}{\mathbf{h}}_{c,k}^H\mathbf{w}_{c,i}x_{c,i}}
\limits_{\textrm{Other users' interference}} \\
&+ \underbrace{{\sum}_{n=1}^{N} {\mathbf{h}}_{c,k}^H\mathbf{w}_{r,n}x_{r,n}}
\limits_{\textrm{Radar sensing interference}} 
 + n_{c,k},\nonumber
\end{align}
where
$n_{c,k} \sim \mathcal{CN}(0, \sigma_{c,k}^2)  $ is the complex additive white Gaussian noise (AWGN) at user $k$.
Furthermore,
let 
$ \bar{\mathbf{h}}_{c,k} \triangleq [ \mathbf{h}_{c,k}^T,\cdots, \mathbf{h}_{c,k}^T ]^T \in \mathbb{C}^{NK \times 1}  $
and
$ \bar{\mathbf{h}}_{c2,k} \triangleq [ \mathbf{h}_{c,k}^T,\cdots, \mathbf{h}_{c,k}^T ]^T \in \mathbb{C}^{NN \times 1}  $.
The signal ${y}_{c,k}$ can be reformulated as
\begin{align}
{y}_{c,k}
&=
\bar{\mathbf{h}}_{c,k}^H\mathbf{B}_{c,k}\mathbf{w}_{c}x_{c,k}
+{\sum}_{i\neq k}^{K}\bar{\mathbf{h}}_{c,k}^H\mathbf{B}_{c,i}\mathbf{w}_{c}x_{c,i} \\
&+ {\sum}_{n=1}^{N} \bar{\mathbf{h}}_{c2,k}^H\mathbf{B}_{r,n}\mathbf{w}_{r}x_{r,n}
+ n_{c,k},\nonumber
\end{align}
where
the above selection matrices 
$\mathbf{B}_{c,k}\in \mathbb{R}^{NK\times NK}$
and
$\mathbf{B}_{r,n}\in \mathbb{R}^{NN\times NN}$
are respectively given as
\begin{align}
&\mathbf{B}_{c,k} \triangleq \text{diag}(\mathbf{b}_{c,k}),
\mathbf{B}_{r,n} \triangleq \text{diag}(\mathbf{b}_{r,n})
\end{align}
and the index vectors 
$\mathbf{b}_{c,k}$ 
and
$\mathbf{b}_{r,n}$ 
are respectively defined as follows
\begin{align}
&\mathbf{b}_{c,k} \triangleq [\underbrace{0,\cdots,0}
\limits_{(k-1)N}
,\underbrace{1,\cdots,1}
\limits_{N}
,0,\cdots,0]^T \in \mathbb{R}^{NK\times 1},\\
&\mathbf{b}_{r,n} \triangleq [\underbrace{0,\cdots,0}
\limits_{(n-1)N}
,\underbrace{1,\cdots,1}
\limits_{N}
,0,\cdots,0]^T \in \mathbb{R}^{NN\times 1}.
\end{align}

Accordingly, the received SINR of the $k$-th user is given by
\begin{align}
&\text{SINR}_{k}(\mathbf{w}_{c}, \mathbf{w}_{r})\\
&=
\frac{\vert\bar{\mathbf{h}}_{c,k}^H\mathbf{B}_{c,k}\mathbf{w}_{c}  \vert^2}
{
\sum_{i\neq k}^{K}\vert \bar{\mathbf{h}}_{c,k}^H\mathbf{B}_{c,i}\mathbf{w}_{c}\vert^2
+
\sum_{n=1}^{N} \vert \bar{\mathbf{h}}_{c2,k}^H\mathbf{B}_{r,n}\mathbf{w}_{r}\vert^2
+
\sigma_{c,k}^2
}.\nonumber
\end{align}
Therefore, 
the achievable rate of the $k$-th user is given as
\begin{align}
\mathrm{R}_{k}( \mathbf{w}_{c}, \mathbf{w}_{r}  ) =\text{log}( 1 + \text{SINR}_{k}( \mathbf{w}_{c}, \mathbf{w}_{r}  )),
\forall k \in \mathcal{K}.
\end{align}

\textit{2) Radar Sensing Model:}
We consider that the channel between the TRIS transceiver and the target is  the line-of-sight (LoS) link.
First,
we define $N \triangleq N_h \times N_v$,
where 
$N_h$ and $N_v$ are the numbers of the TRIS transceiver units in the horizontal and vertical directions, respectively.
The steering vector from the TRIS transceiver to the target can be expressed in (\ref{h_LoS}),
\begin{figure*}
\begin{align}
\mathbf{h}_{r}
=&\alpha_{r}
\bigg[
1,
e^{-j \frac{2\pi}{\lambda} d \sin \theta_{r}^{AoD} \cos\psi_{r}^{AoD}  },
\cdots,
e^{-j \frac{2\pi}{\lambda} (N_h-1) d \sin \theta_{r}^{AoD} \cos\psi_{r}^{AoD}  }
\bigg]^T \label{h_LoS}\\
&\otimes\bigg[
1,
e^{-j \frac{2\pi}{\lambda} d \cos \theta_{r}^{AoD}  },
\cdots,
e^{-j \frac{2\pi}{\lambda} (N_v-1) d \cos \theta_{r}^{AoD} }
\bigg]^T,\nonumber
\end{align}
\boldsymbol{\hrule}
\end{figure*}
where
$\alpha_{r}$ is the
large-scale fading coefficient,
$\theta_{r}^{AoD}$
and
$\psi_{r}^{AoD}$
are the zenith and azimuth angles-of-departure (AoDs) at the TRIS transceiver, respectively,
$d$ denotes the spacing between two adjacent  units of the TRIS transceiver,
and $\lambda$ represents the carrier wavelength.

{
Consequently,
the beampattern gain directed at the target \cite{ref_beampattern} is represented by
{\small
\begin{align}
&\mathrm{P}_{target}
=\mathbb{E}\bigg\{
\bigg\vert  {\sum}_{i=1}^{N} {\mathbf{h}}_{r}^H\mathbf{w}_{r,i}x_{r,i} 
+
{\sum}_{k=1}^{K}{\mathbf{h}}_{r}^H\mathbf{w}_{c,k}x_{c,k}  \bigg\vert^2
\bigg\}
\\
&=\mathbb{E}\bigg\{
\bigg\vert 
\sum_{i=1}^{N}\bar{\mathbf{h}}_{r}^H \mathbf{B}_{r,i}  \mathbf{w}_{r}x_{r,i}
+\sum_{k=1}^{K}\bar{\mathbf{h}}_{r2}^H\mathbf{B}_{c,k} \mathbf{w}_{c}x_{c,k}
\bigg\vert^2
\bigg\}\nonumber\\
&=\mathbf{w}_{r}^H\! \bigg( \! \sum_{i=1}^{N} \mathbf{B}_{r,i}\bar{\mathbf{h}}_{r} \bar{\mathbf{h}}_{r}^H \mathbf{B}_{r,i}  \!\bigg)\mathbf{w}_{r}
\!\!+\!\!\mathbf{w}_{c}^H\! \bigg( \! \sum_{k=1}^{K} \mathbf{B}_{c,k}\bar{\mathbf{h}}_{r2} \bar{\mathbf{h}}_{r2}^H \mathbf{B}_{c,k} \! \bigg)\!\mathbf{w}_{c},\nonumber
\end{align}
}
\par\noindent
where
the newly introduced channel coefficients
$ \bar{\mathbf{h}}_{r} $
and
$ \bar{\mathbf{h}}_{r2} $
are respectively formulated as
\begin{align}
&\bar{\mathbf{h}}_{r} \triangleq [ \mathbf{h}_{r}^T,\cdots, \mathbf{h}_{r}^T ]^T \in \mathbb{C}^{NN \times 1},\\
&\bar{\mathbf{h}}_{r2} \triangleq [ \mathbf{h}_{r}^T,\cdots, \mathbf{h}_{r}^T ]^T \in \mathbb{C}^{NK \times 1}.
\end{align}}

\subsection{Problem Formulation}

\textit{1) Sum-Rate Maximization:}
{The first objective of this paper is to maximize the sum-rate of all mobile users by optimizing the transmit beamforming of the TRIS transceiver,
subject to the target beampattern gain requirement and per-element transmit power limits of the TRIS transceiver.}
Accordingly,
the associated optimization problem can be mathematically written as
\begin{subequations}
\begin{align}
\textrm{(P0)}:&\mathop{\textrm{max}}
\limits_{\mathbf{w}_c, \mathbf{w}_r
}\
{\sum}_{k=1}^{K}
\omega_{c,k}
\mathrm{R}_{k}(\mathbf{w}_c, \mathbf{w}_r) 
\label{P0_obj}\\
\textrm{s.t.}\ 
&  \mathrm{P}_{target}(\mathbf{w}_c, \mathbf{w}_r)  \geq P_r,\label{P0_c_0}\\
& \mathbf{w}_{c}^H \bar{\mathbf{A}}_{c,n} \mathbf{w}_{c} 
+ \mathbf{w}_{r}^H \bar{\mathbf{A}}_{r,n}\mathbf{w}_{r} \leq P_t, \forall n \in \mathcal{N}, \label{P0_c_1}
\end{align}
\end{subequations}
where $\omega_{c,k}$ denotes the weight of the $k$-th user. 
{The ISAC system designer can adjust these weights to account for different user priorities. 
In this paper, 
equal weights are adopted to treat all users fairly and simplify the presentation, i.e., 
$\omega_{c,k}=1,~\forall k \in \mathcal{K}$. 
This setting does not affect the subsequent algorithm design.
$P_r$ in (\ref{P0_c_0}) denotes the minimum beampattern gain of the  target,
$P_t$ in (\ref{P0_c_1}) is the maximum power at the TRIS transceiver units.}

\textit{2) Beampattern Gain Maximization:}
The second objective of this paper is to maximize the beampattern gain at the target,
and the corresponding optimization problem is mathematically formulated as
\begin{subequations}
\begin{align}
\textrm{(P1)}:&\mathop{\textrm{max}}
\limits_{\mathbf{w}_c, \mathbf{w}_r
}\
\mathrm{P}_{target}(\mathbf{w}_c, \mathbf{w}_r)
\label{P1_obj}\\
\textrm{s.t.}\ 
& \mathrm{R}_{k}(\mathbf{w}_c, \mathbf{w}_r) \geq R_{th}, \forall k \in \mathcal{K},\label{P1_c_0}\\
& \mathbf{w}_{c}^H \bar{\mathbf{A}}_{c,n} \mathbf{w}_{c} 
+ \mathbf{w}_{r}^H \bar{\mathbf{A}}_{r,n}\mathbf{w}_{r} \leq P_t, \forall n \in \mathcal{N}, \label{P1_c_1}
\end{align}
\end{subequations}
where 
the constant
$R_{th}$ is the minimum rate requirement for each user.

{
The problems (P0) and (P1) are non-convex due to the objective function and constraints.
In the next sections,
we will solve the above optimization problems in turn.}

\section{Solution To Sum-Rate Maximization Problem}

{In this section, 
we propose an SOCP-based algorithm and a low-complexity ADMM-based algorithm
to solve problem (P0).}

\subsection{Problem Reformulation}

{To make problem (P0) more tractable,
by leveraging the fractional programming (FP) framework \cite{ref_FP},
we will equivalently rewrite the rate function (\ref{P0_obj}).}
Specifically,
by introducing auxiliary variables 
$\{ \gamma_{k} \}$ 
and using the Lagrangian dual reformulation,
the original rate function in (\ref{P0_obj}) can be transformed into (\ref{FP_1}).
\begin{figure*}
\begin{align}
&{\mathrm{\dot{R}}_{k}(\mathbf{w}_c, \mathbf{w}_r, \gamma_{k})
=\text{log}(1+ \gamma_{k})
-
\gamma_{k}
+
\frac{(1+\gamma_{k})\vert\bar{\mathbf{h}}_{c,k}^H\mathbf{B}_{c,k}\mathbf{w}_{c}  \vert^2}
{
\sum_{i=1}^{K}\vert \bar{\mathbf{h}}_{c,k}^H\mathbf{B}_{c,i}\mathbf{w}_{c}\vert^2
+
\sum_{n=1}^{N} \vert \bar{\mathbf{h}}_{c2,k}^H\mathbf{B}_{r,n}\mathbf{w}_{r}\vert^2
+
\sigma_{c,k}^2
}.}\label{FP_1}
\end{align}
\boldsymbol{\hrule}
\end{figure*}
{Furthermore,
by introducing the auxiliary variables $\{  \omega_{k} \}$ and applying the quadratic transform,
the expression in (\ref{FP_1}) can be further converted to (\ref{FP_2}).}
\begin{figure*}
\begin{align}
{\mathrm{\ddot{R}}_{k}(\mathbf{w}_c, \mathbf{w}_r, \gamma_{k}, \omega_{k})}
&{=\text{log}(1+ \gamma_{k})
-
\gamma_{k}
+
\big(
2\sqrt{(1+\gamma_{k})}\text{Re}
\{
\omega_{k}^{\ast}
\bar{\mathbf{h}}_{c,k}^H\mathbf{B}_{c,k}\mathbf{w}_{c}
\}}\label{FP_2}\\
&{-
\vert \omega_{k}\vert^2
( {\sum}_{i=1}^{K}\vert \bar{\mathbf{h}}_{c,k}^H\mathbf{B}_{c,i}\mathbf{w}_{c}\vert^2
+
{\sum}_{n=1}^{N} \vert \bar{\mathbf{h}}_{c2,k}^H\mathbf{B}_{r,n}\mathbf{w}_{r}\vert^2
+
\sigma_{c,k}^2 )
\big).}
\nonumber
\end{align}
\boldsymbol{\hrule}
\end{figure*}

Based on the above transformations, 
problem (P0) can be equivalently formulated as
\begin{subequations}
\begin{align}
\textrm{(P2)}:&\mathop{\textrm{max}}
\limits_{\mathbf{w}_c, \mathbf{w}_r, 
\{ \gamma_{k} \},
\{  \omega_{k} \}
}\
{\sum}_{k=1}^{K}
\mathrm{\ddot{R}}_{k}(\mathbf{w}_c, \mathbf{w}_r, \gamma_{k}, \omega_{k}) 
\label{P2_obj}\\
\textrm{s.t.}\ 
&  \mathrm{P}_{target}(\mathbf{w}_c, \mathbf{w}_r)  \geq P_r,\label{P2_c_0}\\
& \mathbf{w}_{c}^H \bar{\mathbf{A}}_{c,n} \mathbf{w}_{c} 
+ \mathbf{w}_{r}^H \bar{\mathbf{A}}_{r,n}\mathbf{w}_{r} \leq P_t, \forall n \in \mathcal{N}. \label{P2_c_1}
\end{align}
\end{subequations}

Observe that the beamforming variables $\{\mathbf{w}_c, \mathbf{w}_r\}$ 
and the auxiliary variables $\{ \gamma_{k}, \omega_{k} \}$
are coupled in the objective function (\ref{P2_obj}),
making problem (P2) difficult to solve directly.

Thus,
the block coordinate ascent (BCA) method \cite{ref_BCA} is implemented to tackle problem (P2).
Specifically,
we divide the optimization variables in problem (P2) into three blocks, 
i.e., 
$\{\mathbf{w}_c, \mathbf{w}_r\}$,
$\{ \gamma_{k} \}$,
and
$\{  \omega_{k} \}$.
{The objective of (P2) is maximized via a blockwise alternating optimization scheme: 
at each iteration one block is optimized while the others are held fixed, 
and the process is repeated until convergence.}

\subsection{Optimizing Auxiliary Variables}

\textit{1) Update $\{ \gamma_{k} \}$:}
When the other variables are given,
the problem of optimizing the auxiliary variable $ \gamma_{k} $ is an unconstrained optimization problem.
The optimal solution $\gamma_{k}^{\star}$ 
is obtained by solving
$\frac{\partial \mathrm{\ddot{R}}_{k}(\mathbf{w}_c, \mathbf{w}_r, \gamma_{k}, \omega_{k})}{\partial \gamma_{k}} = 0$.
Hence,
the optimal solution $\gamma_{k}^{\star}$ can be expressed as
\begin{align}
\gamma_{k}^{\star}
\!\!=\!\!
\frac{\vert\bar{\mathbf{h}}_{c,k}^H\mathbf{B}_{c,k}\mathbf{w}_{c}  \vert^2}
{
\sum_{i\neq k}^{K}\vert \bar{\mathbf{h}}_{c,k}^H\mathbf{B}_{c,i}\mathbf{w}_{c}\vert^2
\!\!+\!\!
\sum_{n=1}^{N} \vert \bar{\mathbf{h}}_{c2,k}^H\mathbf{B}_{r,n}\mathbf{w}_{r}\vert^2
\!\!+\!\!
\sigma_{c,k}^2
}.\label{FP_1_solution}
\end{align}

\textit{2) Update $\{ \omega_{k} \}$:}
Similarly,
the optimal solution $\omega_{k}^{\star}$ can be easily obtained by setting 
$\frac{\partial \mathrm{\ddot{R}}_{k}(\mathbf{w}_c, \mathbf{w}_r, \gamma_{k}, \omega_{k})}{\partial \omega_{k}^{\ast}} = 0$,
which can be formulated as
\begin{align}
\omega_{k}^{\star}
\!\!=\!\!\frac{\sqrt{1+\gamma_{k}}\bar{\mathbf{h}}_{c,k}^H\mathbf{B}_{c,k}\mathbf{w}_{c}}
{
\sum_{i=1}^{K}\vert \bar{\mathbf{h}}_{c,k}^H\mathbf{B}_{c,i}\mathbf{w}_{c}\vert^2
\!\!+\!\!
\sum_{n=1}^{N} \vert \bar{\mathbf{h}}_{c2,k}^H\mathbf{B}_{r,n}\mathbf{w}_{r}\vert^2
\!\!+\!\!
\sigma_{c,k}^2
}.\label{FP_2_solution}
\end{align}

\subsection{Updating the Beamformers}
In this subsection, 
we study the optimization of the transmit beamformers $\{\mathbf{w}_c, \mathbf{w}_r\}$.
First, we introduce the following notations
\begin{align}
&\mathbf{b}_{1,k}
\triangleq
\sqrt{1+\gamma_{k}}(\omega_{k}\mathbf{B}_{c,k}\bar{\mathbf{h}}_{c,k}),\\
&c_{1,k}
\triangleq
\text{log}(1+\gamma_{k})-\gamma_{k}-\vert\omega_{k}\vert^2\sigma_{c,k}^2,\nonumber\\
&\mathbf{B}_{1,k}
\triangleq
{\sum}_{i=1}^{K}\vert\omega_{k}\vert^2\mathbf{B}_{c,i}\bar{\mathbf{h}}_{c,k}\bar{\mathbf{h}}_{c,k}^H\mathbf{B}_{c,i},\nonumber\\
&\mathbf{B}_{2,k}
\triangleq
{\sum}_{n=1}^{N}\vert\omega_{k}\vert^2\mathbf{B}_{r,n}\bar{\mathbf{h}}_{c2,k}\bar{\mathbf{h}}_{c2,k}^H\mathbf{B}_{r,n}.\nonumber
\end{align}

{
Following the above notations, 
we can rearrange the function $\mathrm{\ddot{R}}_{k}(\mathbf{w}_c, \mathbf{w}_r)$ into 
a function with respect to (w.r.t.) the variables $\{\mathbf{w}_c, \mathbf{w}_r\}$ explicitly, 
which is formulated as
\begin{align}
&\mathrm{\ddot{R}}_{k}(\mathbf{w}_c, \mathbf{w}_r) \label{Obj_trans}\\
&=
- \mathbf{w}_c^H \mathbf{B}_{1,k}\mathbf{w}_c - \mathbf{w}_r^H\mathbf{B}_{2,k}\mathbf{w}_r
+
2\text{Re}\{ \mathbf{b}_{1,k}^H\mathbf{w}_c \} + c_{1,k}.\nonumber
\end{align}
Thus,
the objective (\ref{P2_obj}) can be rewritten as follows
\begin{align}
&{\sum}_{k=1}^{K} \mathrm{\ddot{R}}_{k}(\mathbf{w}_c, \mathbf{w}_r) \\
&=
- \mathbf{w}_c^H \mathbf{B}_{3}\mathbf{w}_c - \mathbf{w}_r^H\mathbf{B}_{4}\mathbf{w}_r
+
2\text{Re}\{ \mathbf{b}_{2}^H\mathbf{w}_c \} + c_{2},\nonumber
\end{align}}
where
\begin{align}
&\mathbf{B}_{3} \triangleq {\sum}_{k=1}^{K}\mathbf{B}_{1,k},
\mathbf{B}_{4} \triangleq {\sum}_{k=1}^{K}\mathbf{B}_{2,k},\\
&\mathbf{b}_{2}\triangleq {\sum}_{k=1}^{K}\mathbf{b}_{1,k},
c_{2}\triangleq {\sum}_{k=1}^{K}\mathbf{c}_{1,k}.\nonumber
\end{align}

{
At the same time, we can introduce 
$\mathbf{B}_{5}
\triangleq
{\sum}_{i=1}^{K}\mathbf{B}_{c,i}\bar{\mathbf{h}}_{r2}\bar{\mathbf{h}}_{r2}^H\mathbf{B}_{c,i}$
and
$
\mathbf{B}_{6}
\triangleq
{\sum}_{n=1}^{N}\mathbf{B}_{r,n}\bar{\mathbf{h}}_{r}\bar{\mathbf{h}}_{r}^H\mathbf{B}_{r,n}$.}
The beampattern gain constraint (\ref{P2_c_0}) can be written as
\begin{align}
\mathrm{P}_{target}(\mathbf{w}_c, \mathbf{w}_r)
=
\mathbf{w}_c^H \mathbf{B}_{5}\mathbf{w}_c + \mathbf{w}_r^H\mathbf{B}_{6}\mathbf{w}_r.\label{beampattern_trans}
\end{align}

Therefore, 
based on the above transformations, 
the beamforming optimization problem is given as follows
\begin{subequations}
\begin{align}
\textrm{(P3)}:&\mathop{\textrm{max}}
\limits_{\mathbf{w}_c, \mathbf{w}_r
}\
\!-\! \mathbf{w}_c^H \mathbf{B}_{3}\mathbf{w}_c 
\!\!-\!\! \mathbf{w}_r^H\mathbf{B}_{4}\mathbf{w}_r
\!\!+\!\!
2\text{Re}\{ \mathbf{b}_{2}^H\mathbf{w}_c \} 
\!\!+\!\! c_{2}
\label{P3_obj}\\
\textrm{s.t.}\ 
&  \mathbf{w}_c^H \mathbf{B}_{5}\mathbf{w}_c + \mathbf{w}_r^H\mathbf{B}_{6}\mathbf{w}_r  \geq P_r,\label{P3_c_0}\\
& \mathbf{w}_{c}^H \bar{\mathbf{A}}_{c,n} \mathbf{w}_{c} 
+ \mathbf{w}_{r}^H \bar{\mathbf{A}}_{r,n}\mathbf{w}_{r} \leq P_t, \forall n \in \mathcal{N}. \label{P3_c_1}
\end{align}
\end{subequations}

It is observed that the radar beampattern gain constraint (\ref{P3_c_0}) is non-convex.
To solve it,
the MM framework is applied.
Specifically,
inspired by the MM methodology \cite{ref_MM},
we can obtain a tight lower bound of the term 
$\mathbf{w}_c^H \mathbf{B}_{5}\mathbf{w}_c$ in (\ref{P3_c_0})
at the point $\mathbf{w}_{c,0}$,
which is given as
\begin{align}
\mathbf{w}_c^H \mathbf{B}_{5}\mathbf{w}_c
\!\geq\!
\mathbf{w}_{c,0}^H \mathbf{B}_{5}\mathbf{w}_{c,0}
+
2\text{Re}\{ \mathbf{w}_{c,0}^H \mathbf{B}_{5}(\mathbf{w}_{c} - \mathbf{w}_{c,0}  )  \},\label{SOCP_MM_1}
\end{align}
where $\mathbf{w}_{c,0}$ is obtained from the previous iteration.
Similarly,
the tight lower bound of the term 
$\mathbf{w}_r^H \mathbf{B}_{6}\mathbf{w}_r$ in (\ref{P3_c_0})
can be obtained as
\begin{align}
\mathbf{w}_r^H \mathbf{B}_{6}\mathbf{w}_r
\!\geq\!
\mathbf{w}_{r,0}^H \mathbf{B}_{6}\mathbf{w}_{r,0}
+
2\text{Re}\{ \mathbf{w}_{r,0}^H \mathbf{B}_{6}(\mathbf{w}_{r} - \mathbf{w}_{r,0}  )  \},\label{SOCP_MM_2}
\end{align}
where
$\mathbf{w}_{r,0}$ is the value obtained in the last iteration.

As a result,
by replacing the quadratic terms
$\mathbf{w}_c^H \mathbf{B}_{5}\mathbf{w}_c$
and
$\mathbf{w}_r^H \mathbf{B}_{6}\mathbf{w}_r$ by
(\ref{SOCP_MM_1})
and
(\ref{SOCP_MM_2}),
respectively,
problem (P3) is reformulated as
\begin{subequations}
\begin{align}
\textrm{(P4)}:&\mathop{\textrm{max}}
\limits_{\mathbf{w}_c, \mathbf{w}_r
}\
\!-\! \mathbf{w}_c^H \mathbf{B}_{3}\mathbf{w}_c 
\!\!-\!\! \mathbf{w}_r^H\mathbf{B}_{4}\mathbf{w}_r
\!\!+\!\!
2\text{Re}\{ \mathbf{b}_{2}^H\mathbf{w}_c \} 
\!\!+\!\! c_{2}
\label{P4_obj}\\
\textrm{s.t.}\ 
&  \mathbf{w}_{c,0}^H \mathbf{B}_{5}\mathbf{w}_{c,0}
+
2\text{Re}\{ \mathbf{w}_{c,0}^H \mathbf{B}_{5}(\mathbf{w}_{c} - \mathbf{w}_{c,0}  )  \}\label{P4_c_0}\\
& + \mathbf{w}_{r,0}^H \mathbf{B}_{6}\mathbf{w}_{r,0}
+
2\text{Re}\{ \mathbf{w}_{r,0}^H \mathbf{B}_{6}(\mathbf{w}_{r} - \mathbf{w}_{r,0}  )  \}  \geq P_r,\nonumber\\
& \mathbf{w}_{c}^H \bar{\mathbf{A}}_{c,n} \mathbf{w}_{c} 
+ \mathbf{w}_{r}^H \bar{\mathbf{A}}_{r,n}\mathbf{w}_{r} \leq P_t, \forall n \in \mathcal{N}. \label{P4_c_1}
\end{align}
\end{subequations}

{Problem (P4) is a typical SOCP and can be solved via off-the-shelf numerical solvers, e.g., CVX \cite{ref_CVX}. 
The proposed SOCP-based algorithm is given in Alg. \ref{alg:1}.}
\begin{algorithm}[t]
\caption{Solving the Problem (P0)}
\label{alg:1}
\begin{algorithmic}[1]
\STATE {initialize}
$\mathbf{w}_c^{(0)}$,
$\mathbf{w}_r^{(0)}$,
and
$t=0$;
\REPEAT
\STATE update $\{\gamma_k^{(t+1)}\}$ and $\{\omega_k^{(t+1)}\}$ by (\ref{FP_1_solution}) and (\ref{FP_2_solution}), respectively;
\STATE update $\mathbf{w}_c^{(t+1)}$ and $\mathbf{w}_r^{(t+1)}$ by solving  (P4);
\STATE $t++$;
\UNTIL{$convergence$;}
\end{algorithmic}
\end{algorithm}

\subsection{Low-Complexity Solution}

However,
since general convex optimization solvers like CVX rely on the interior-point (IP) method \cite{ref_Convex Optimization} for the SOCP problem, 
the complexity of solving the SOCP problem rises sharply with the variable dimension and can be prohibitively high.
This difficulty motivates us to explore a more efficient solution that avoids reliance on numerical optimization solvers.

First, 
let $\mathbf{w}_{cr}
\triangleq
[\mathbf{w}_{c}^T,\mathbf{w}_{r}^T  ]^T$.
Then, 
the objective function in (\ref{P4_obj}) 
can be reformulated into the following compact form
\begin{align}
&- \mathbf{w}_c^H \mathbf{B}_{3}\mathbf{w}_c - \mathbf{w}_r^H\mathbf{B}_{4}\mathbf{w}_r
+
2\text{Re}\{ \mathbf{b}_{2}^H\mathbf{w}_c \} + c_{2}\\
&\Leftrightarrow
- \mathbf{w}_{cr}^H \mathbf{B}_{7}\mathbf{w}_{cr}
+
2\text{Re}\{ \mathbf{b}_{3}^H\mathbf{w}_{cr} \} + c_{2},\nonumber
\end{align}
where
$\mathbf{b}_{3}\triangleq[\mathbf{b}_{2}^T,\mathbf{0}^T  ]^T$
and
$\mathbf{B}_7 \triangleq \text{blkdiag}( \mathbf{B}_3, \mathbf{B}_4)$.

Furthermore,
the radar beampattern gain constraint (\ref{P4_c_0}) can be reformulated as
\begin{align}
&{  \mathbf{w}_{c,0}^H \mathbf{B}_{5}\mathbf{w}_{c,0}
+
2\text{Re}\{ \mathbf{w}_{c,0}^H \mathbf{B}_{5}(\mathbf{w}_{c} - \mathbf{w}_{c,0}  )  \}}\\
&{ + \mathbf{w}_{r,0}^H \mathbf{B}_{6}\mathbf{w}_{r,0}
+
2\text{Re}\{ \mathbf{w}_{r,0}^H \mathbf{B}_{6}(\mathbf{w}_{r} - \mathbf{w}_{r,0}  )  \}\geq P_r}\nonumber\\
&{\Leftrightarrow
- 2\text{Re}\{ \mathbf{b}_{4}^H \mathbf{w}_{cr} \} - c_{3} \leq 0,} \nonumber
\nonumber
\end{align}
where
\begin{align}
&\mathbf{B}_{8} \triangleq \text{blkdiag}( \mathbf{B}_5, \mathbf{B}_6),  
\mathbf{w}_{cr,0}\triangleq [\mathbf{w}_{c,0}^T,\mathbf{w}_{r,0}^T  ]^T,\\
&\mathbf{b}_{4} \triangleq \mathbf{B}_{8}^H\mathbf{w}_{cr,0},
c_3 \triangleq -( \mathbf{w}_{cr,0}^H \mathbf{B}_{8} \mathbf{w}_{cr,0} )^{\ast}-P_r.\nonumber
\end{align}

Similarly,
the per-unit power constraints (\ref{P4_c_1}) can be rewritten as
\begin{align}
&  \mathbf{w}_{c}^H \bar{\mathbf{A}}_{c,n} \mathbf{w}_{c} 
+ \mathbf{w}_{r}^H \bar{\mathbf{A}}_{r,n}\mathbf{w}_{r} \leq P_t\\
&\Leftrightarrow
\mathbf{w}_{cr}^H \bar{\mathbf{A}}_{cr,n} \mathbf{w}_{cr} \leq P_t, \nonumber
\nonumber
\end{align}
where 
$\bar{\mathbf{A}}_{cr,n}\triangleq \text{blkdiag}(\bar{\mathbf{A}}_{c,n}, \bar{\mathbf{A}}_{r,n} ) $.

With the above transformations, 
the optimization problem (P4) is transformed into the following formulation
\begin{subequations}
\begin{align}
\textrm{(P5)}:&\mathop{\textrm{min}}
\limits_{\mathbf{w}_{cr}
}\
 \mathbf{w}_{cr}^H \mathbf{B}_{7}\mathbf{w}_{cr}
-
2\text{Re}\{ \mathbf{b}_{3}^H\mathbf{w}_{cr} \} - c_{2}
\label{P5_obj}\\
\textrm{s.t.}\ 
&- 2\text{Re}\{ \mathbf{b}_{4}^H \mathbf{w}_{cr} \} - c_{3} \leq 0,\label{P5_c_0}\\
&\mathbf{w}_{cr}^H \bar{\mathbf{A}}_{cr,n} \mathbf{w}_{cr}  \leq P_t, \forall n \in \mathcal{N}. \label{P5_c_1}
\end{align}
\end{subequations}

By introducing one copy of the variable $\mathbf{w}_{cr}$,
i.e.,
$\mathbf{w}_{cr} = \mathbf{f}$,
problem (P5) can be rewritten as follows
\begin{subequations}
\begin{align}
\textrm{(P6)}:&\mathop{\textrm{min}}
\limits_{\mathbf{w}_{cr}, \mathbf{f}
}\
 \mathbf{w}_{cr}^H \mathbf{B}_{7}\mathbf{w}_{cr}
-
2\text{Re}\{ \mathbf{b}_{3}^H\mathbf{w}_{cr} \} - c_{2}
\label{P6_obj}\\
\textrm{s.t.}\ 
&- 2\text{Re}\{ \mathbf{b}_{4}^H \mathbf{w}_{cr} \} - c_{3} \leq 0,\label{P6_c_0}\\
&\mathbf{f}^H \bar{\mathbf{A}}_{cr,n} \mathbf{f}  \leq P_t, \forall n \in \mathcal{N}, \label{P6_c_1}\\
&\mathbf{w}_{cr} = \mathbf{f}.\label{P6_c_2}
\end{align}
\end{subequations}

Next,
we adopt the ADMM methodology \cite{ref_ADMM} to solve problem (P6).
By relaxing the equality constraint (\ref{P6_c_2}) and penalizing it in the objective, 
the augmented Lagrangian (AL) function of (P6) is formulated by
\begin{align}
\mathcal{L}( \mathbf{w}_{cr}, \mathbf{f}, \boldsymbol{\tau} )
&=
 \mathbf{w}_{cr}^H \mathbf{B}_{7}\mathbf{w}_{cr}
-
2\text{Re}\{ \mathbf{b}_{3}^H\mathbf{w}_{cr} \} - c_{2}\\
&+\text{Re}\{ \boldsymbol{\tau}^H( \mathbf{w}_{cr} - \mathbf{f} ) \}
+\frac{\rho}{2}\Vert\mathbf{w}_{cr} - \mathbf{f}\Vert_2^2,\nonumber
\end{align}
where $\rho$ is a positive constant
and 
$\boldsymbol{\tau} \in \mathbb{C}^{N(K+N) \times 1}$ is the Lagrangian multiplier associated with the constraint (\ref{P6_c_2}).
Therefore,
problem (P6) is reformulated as
\begin{subequations}
\begin{align}
\textrm{(P7)}:&\mathop{\textrm{min}}
\limits_{\mathbf{w}_{cr}, \mathbf{f}, \boldsymbol{\tau}
}\
\mathcal{L}( \mathbf{w}_{cr}, \mathbf{f}, \boldsymbol{\tau} )
\label{P7_obj}\\
\textrm{s.t.}\ 
&- 2\text{Re}\{ \mathbf{b}_{4}^H \mathbf{w}_{cr} \} - c_{3} \leq 0,\label{P7_c_0}\\
&\mathbf{f}^H \bar{\mathbf{A}}_{cr,n} \mathbf{f}  \leq P_t, \forall n \in \mathcal{N}.\label{P7_c_1}
\end{align}
\end{subequations}

{
Subsequently, 
following the ADMM framework,
problem (P7) can be solved by separately updating the variables 
$\mathbf{w}_{cr}$, 
$\mathbf{f}$,
and 
$\boldsymbol{\tau}$
using the block coordinate descent (BCD) method, as shown below:
\begin{subequations}
\begin{align}
&\mathbf{w}_{cr}^{(t+1)}= \mathop{\text{arg min}}
\limits_{- 2\text{Re}\{ \mathbf{b}_{4}^H \mathbf{w}_{cr} \} - c_{3} \leq 0}
\mathcal{L}( \mathbf{w}_{cr}, \mathbf{f}^{(t)}, \boldsymbol{\tau}^{(t)} ),\\
&\mathbf{f}^{(t+1)}= \mathop{\text{arg min}}
\limits_{\mathbf{f}^H \bar{\mathbf{A}}_{cr,n} \mathbf{f}  \leq P_t, \forall n \in \mathcal{N}}
\mathcal{L}( \mathbf{w}_{cr}^{(t+1)}, \mathbf{f}, \boldsymbol{\tau}^{(t)} ),\\
&\boldsymbol{\tau}^{(t+1)}= \boldsymbol{\tau}^{(t)} + \rho(\mathbf{w}_{cr}^{(t+1)}-\mathbf{f}^{(t+1)}).
\end{align}
\end{subequations}}

The update procedures are detailed as follows.

\underline{\textit{1) Update $\mathbf{w}_{cr}$:}}
With $\mathbf{f}$ and $\boldsymbol{\tau}$ being fixed, 
the minimization of $\mathcal{L}( \mathbf{w}_{cr}, \mathbf{f}, \boldsymbol{\tau} )$
w.r.t. $\mathbf{w}_{cr}$ reduces to the following problem
\begin{subequations}
\begin{align}
\textrm{(P8)}:&\mathop{\textrm{min}}
\limits_{ \mathbf{w}_{cr}
}\
\mathbf{w}_{cr}^H \bar{\mathbf{B}}_{7} \mathbf{w}_{cr}
-
2\text{Re}\{ \bar{\mathbf{b}}_{3}^H \mathbf{w}_{cr} \} - \bar{c}_{2}\label{P8_obj}
\\
&\textrm{s.t.}\ 
 - 2\text{Re}\{ \mathbf{b}_{4}^H \mathbf{w}_{cr} \} - c_{3} \leq 0,
\end{align}
\end{subequations}
where
\begin{align}
&\bar{\mathbf{B}}_{7} 
\triangleq 
{\mathbf{B}}_{7} + \frac{\rho}{2}\mathbf{I},
\bar{\mathbf{b}}_{3}
\triangleq 
\mathbf{b}_{3} - \frac{1}{2}\boldsymbol{\tau} + \frac{\rho}{2}\mathbf{f},\\
&\bar{c}_{2}
\triangleq 
{c}_{2} + \text{Re}\{ \boldsymbol{\tau}^H\mathbf{f} \} - \frac{\rho}{2}\mathbf{f}^H\mathbf{f}.\nonumber
\end{align}

The above problem (P8) can be efficiently solved by applying the following lemma, which is proved in \cite{ref_QC}.
\begin{lemma}\label{lemma_1}
First, consider the following optimization problem:
\begin{subequations}
\begin{align}
\textrm{(P$_{Lem}$)}:&\mathop{\textrm{min}}
\limits_{ \mathbf{x}
}\
\mathbf{x}^H \mathbf{B} \mathbf{x}
-
 2\text{Re}\{ \mathbf{b}^H \mathbf{x} \} - b\label{P8_obj}
\\
&\textrm{s.t.}\ 
\mathbf{x}^H \bar{\mathbf{B}} \mathbf{x}
 - 2\text{Re}\{ \bar{\mathbf{b}}^H \mathbf{x} \} - \bar{b} \leq 0,
\end{align}
\end{subequations}
where $\mathbf{B} \succ {0}$ and $ \bar{\mathbf{B}}\succeq {0}$, and Slater's condition holds.
The optimal solution of the problem (P$_{Lem}$) can be obtained according to the following two cases:

\begin{itemize}
\item[] \underline{CASE-I}:
 If the inequality $(\mathbf{B}^{-1}\mathbf{b})^H\bar{\mathbf{B}}(\mathbf{B}^{-1}\mathbf{b})
-2\textrm{Re}\{\bar{\mathbf{b}}^H(\mathbf{B}^{-1}\mathbf{b})\}-\bar{b}\leq 0$ holds,
then the optimal solution of $\mathbf{x}$ is given by
 \begin{align}
\mathbf{x}^{\star} 
=
{\mathbf{B}}^{-1}{\mathbf{b}}.
\end{align}

\item[] \underline{CASE-II}:
Otherwise, 
we can obtain
$\mathbf{x}^{\star} = (\mu^{\star}\bar{\mathbf{B}}+\mathbf{B})^{-1}(\mu^{\star}\bar{\mathbf{b}}+\mathbf{b})$,
where $\mu^{\star}$ represents the solution to the following equation
\begin{align}
&\big((\mu^{\star}\bar{\mathbf{B}}+\mathbf{B})^{-1}(\mu^{\star}\bar{\mathbf{b}}+\mathbf{b})\big)^H
\bar{\mathbf{B}}
(\mu^{\star}\bar{\mathbf{B}}+\mathbf{B})^{-1}(\mu^{\star}\bar{\mathbf{b}}+\mathbf{b})\nonumber\\
&-2\textrm{Re}\{\bar{\mathbf{b}}^H(\mu^{\star}\bar{\mathbf{B}}+\mathbf{B})^{-1}
(\mu^{\star}\bar{\mathbf{b}}+\mathbf{b})\}-\bar{b}= 0,
\end{align}
with the value of $\mu^{\star}$  obtained via Newton's method.
\end{itemize}
\end{lemma}

\underline{\textit{2) Update $\mathbf{f}$:}}
In the next step, we explore the update of the variable $\mathbf{f}$, 
which is obtained by minimizing $\mathcal{L}( \mathbf{w}_{cr}, \mathbf{f}, \boldsymbol{\tau} )$. 
With variables $\mathbf{w}_{cr}$ and $\boldsymbol{\tau}$ being fixed, 
this subproblem is reduced to
\begin{subequations}
\begin{align}
\textrm{(P9)}:&\mathop{\textrm{min}}
\limits_{ \mathbf{f}
}\
\frac{\rho}{2}  \mathbf{f}^H \mathbf{f}
-
 2\text{Re}\{ {\mathbf{\hat{b}}}_{3}^H \mathbf{f} \} - \hat{c}_{2}\label{P9_obj}
\\
\textrm{s.t.}\ 
& \mathbf{f}^H\mathbf{\bar{A}}_{cr,n}\mathbf{f} \leq P_t, \forall n \in \mathcal{N},
\end{align}
\end{subequations}
where
\begin{align}
& \hat{c}_{2}
\triangleq 
\mathbf{w}_{cr}^H \mathbf{B}_{7}\mathbf{w}_{cr}
-
 2\text{Re}\{ \mathbf{b}_{3}^H \mathbf{w}_{cr} \} - c_{2}\\
& + \text{Re}\{ \boldsymbol{\tau}^H\mathbf{w}_{cr}   \}
 + \frac{\rho}{2} \Vert\mathbf{w}_{cr}  \Vert_2^2,
{\mathbf{\hat{b}}}_{3}
\triangleq 
\frac{1}{2}
(\boldsymbol{\tau} +\rho \mathbf{w}_{cr} ).\nonumber
\end{align}

Clearly,
the above optimization problem (P9) can be broken down into $N$ independent small problems, which can be solved independently in parallel.
As a result,
each subproblem can be formulated as
\begin{subequations}
\begin{align}
\textrm{(P10)}:&\mathop{\textrm{min}}
\limits_{ \mathbf{f}_n
}\
\mathbf{f}_n^H \mathbf{f}_n
-
 2\text{Re}\{ {\mathbf{b}}_{5,n}^H \mathbf{f}_n \}\label{P10_obj}
\\
\textrm{s.t.}\ 
& \mathbf{f}_n^H\mathbf{f}_n \leq P_t,\label{P10_c_1}
\end{align}
\end{subequations}
where
\begin{align}
& \mathbf{b}_{5} 
\triangleq 
\frac{2}{\rho}
\mathbf{\hat{b}}_{3},\\
& \mathbf{f}_n
\triangleq 
\big[ \mathbf{f}(n),\mathbf{f}(n+N),\cdots, \mathbf{f}(n+(K-1)N), \nonumber\\
&\cdots, \mathbf{f}(n+(K+N-1)N)  \big]^T \in \mathbb{C}^{(K+N)\times 1},\nonumber\\
& \mathbf{b}_{5,n}
\triangleq 
\big[ \mathbf{b}_{5}(n),\mathbf{b}_{5}(n+N),\cdots, \mathbf{b}_{5}(n+(K-1)N),\nonumber \\
&\cdots, \mathbf{b}_{5}(n+(K+N-1)N)  \big]^T \in \mathbb{C}^{(K+N)\times 1}.\nonumber
\end{align}

In the following,
to efficiently solve the above problem (P10), 
we propose a closed-form solution via the Lagrangian multiplier method.
First,
the Lagrange function associated with problem (P10) can be formulated as
\begin{align}
\mathcal{L}(\mathbf{f}_{n},\mu) = \mathbf{f}_n^H \mathbf{f}_n
-
2\text{Re}\{ {\mathbf{b}}_{5,n}^H \mathbf{f}_n \}
+ \mu (\mathbf{f}_n^H\mathbf{f}_n - P_t), \label{P10_1}
\end{align}
where $\mu$ is the Lagrangian multiplier.

Then,
by computing the first-order derivative of the Lagrange function $\mathcal{L}(\mathbf{f}_{n},\mu)$ 
w.r.t. the variable $\mathbf{f}_{n}$ and setting it equal to zero,
we derive the formulation below
\begin{align}
\frac{\partial \mathcal{L}(\mathbf{f}_{n},\mu)}{\partial \mathbf{f}_{n}^{\ast} } = \mathbf{0}.
\end{align}

{
Therefore,
the optimal solution of $\mathbf{f}_{n}$ can be given as
\begin{align}
\mathbf{f}_{n} = \frac{\mathbf{b}_{5,n}}{({ 1 + \mu  })}. \label{P14_closed_solution}
\end{align}

By plugging the optimal expression from (\ref{P14_closed_solution}) into the power constraint (\ref{P10_c_1}), 
we obtain the resulting equation below
\begin{align}
 \frac{\mathbf{b}_{5,n}^H\mathbf{b}_{5,n}}{ (1+\mu )^2  }\leq P_t.  \label{P14_proof_power}
\end{align}

Equation (\ref{P14_proof_power}) reveals that the left-hand side is a monotonically decreasing function of the Lagrangian multiplier $\mu$.
Consequently, 
the optimal solution of problem (P10) falls into one of the following two cases:
\begin{itemize}
\item[] \uwave{CASE-I}:
If $\mu = 0$,
the inequality (\ref{P14_proof_power}) is satisfied.
The optimal solution of (P10) can be formulated as 
\begin{align}
\mathbf{f}_{n}^{\star} = \mathbf{b}_{5,n}. 
\end{align}

\item[] \uwave{CASE-II}: 
Otherwise,
$\mu $ is positive.
And the optimal solution of problem (P10) is represented as
\begin{align}
\mathbf{f}_{n}^{\star} = \sqrt{P_t} \frac{\mathbf{b}_{5,n}}{  \Vert\mathbf{b}_{5,n}\Vert_2  }. \label{P10_2}
\end{align}

\end{itemize}}

\underline{\textit{3) Update $\boldsymbol{\tau}$:}}
According to the ADMM method \cite{ref_ADMM},
after performing the primal variable updates,
the dual variable $\boldsymbol{\tau}$ can be updated by leveraging the gradient ascent method,
which is expressed as
\begin{align}
\boldsymbol{\tau}^{(t+1)} := \boldsymbol{\tau}^{(t)} + \rho(\mathbf{w}_{cr}^{(t+1)}-\mathbf{f}^{(t+1)}).\label{ADMM_rho}
\end{align}

The ADMM-based solution for solving problem (P5) 
is summarized in Alg. \ref{alg:2}.

\begin{algorithm}[t]
\caption{The ADMM-Based Method to Solve (P5)}
\label{alg:2}
\begin{algorithmic}[1]
\STATE {initialize}
$\mathbf{w}_{cr}^{(0)}$,
$\mathbf{f}^{(0)}$,
$\boldsymbol{\tau}^{(0)}$,
and
$t=0$;
\REPEAT
\STATE update $\mathbf{w}_{cr}^{(t+1)}$ by solving (P8);
\FOR{ $n = 1:N$ }
\STATE update $\mathbf{f}_{n}^{(t+1)}$ by solving (P10);
\ENDFOR
\STATE update $\boldsymbol{\tau}^{(t+1)}$ by  (\ref{ADMM_rho});
\STATE $t++$;
\UNTIL{$convergence$;}
\end{algorithmic}
\end{algorithm}

\subsection{Convergence and Complexity Analysis}

\textit{1) Convergence Analysis:}
{
Since Alg. \ref{alg:1} follows the MM methodology \cite{ref_MM}, \cite{ref_convergence}, 
the original objective value of problem (P0) is monotonically increasing after each variable block update. 
According to \cite{ref_convergence}, 
any limit point of the iterates generated by Alg. \ref{alg:1} is a stationary point of the original problem (P0).}

{
Besides, the convergence of the ADMM-based algorithm (i.e., Alg. \ref{alg:2}) follows from the standard convergence results of ADMM-type algorithms \cite{ref_ADMM}.}

\textit{2) Complexity Analysis:}
We next analyze the computational complexity of the proposed algorithms. 
According to \cite{ref_complexity}, 
the per-iteration computational cost of solving problem (P4) is 
$\mathcal{O}(N^{3.5}(K+N)^3)$. 
Accordingly, 
the total complexity of Algorithm~1 can be expressed as
$
\mathcal{O}_{\rm Alg1} = \mathcal{O}(C_1 \, N^{3.5}(K+N)^3 )
$,
where $C_1$ denotes the number of iterations required for convergence of (P0).
For the proposed ADMM-based algorithm, 
the complexity of solving problem (P5)
scales as 
$
\mathcal{O}_{\rm ADMM} = \mathcal{O}(C_2 \, N^{3}(K+N)^3 )
$,
where $C_2$ is the number of ADMM iterations.

\section{Solution To Beampattern Gain Maximization Problem}

In this section, 
we investigate the maximization of the beampattern gain at the target.

\subsection{Problem Reformulation}
To solve problem (P1),
following similar arguments used in the sum-rate maximization problem,
we again use the FP method to transform the rate function $\mathrm{R}_{k}(\mathbf{w}_c, \mathbf{w}_r)$ in constraints (\ref{P1_c_0}),
and then problem (P1) is rewritten as 
\begin{subequations}
\begin{align}
\textrm{(P11)}:&\mathop{\textrm{max}}
\limits_{\mathbf{w}_c, \mathbf{w}_r,
\{\gamma_{k}\}, \{\omega_{k}\}
}\
\mathrm{P}_{target}(\mathbf{w}_c, \mathbf{w}_r)
\label{P11_obj}\\
\textrm{s.t.}\ 
& \mathrm{\ddot{R}}_{k}(\mathbf{w}_c, \mathbf{w}_r, \gamma_{k}, \omega_{k}) \geq R_{th}, \forall k \in \mathcal{K},\label{P11_c_0}\\
& \mathbf{w}_{c}^H \bar{\mathbf{A}}_{c,n} \mathbf{w}_{c} 
+ \mathbf{w}_{r}^H \bar{\mathbf{A}}_{r,n}\mathbf{w}_{r} \leq P_t, \forall n \in \mathcal{N}, \label{P11_c_1}
\end{align}
\end{subequations}
where
the function $\mathrm{\ddot{R}}_{k}(\mathbf{w}_c, \mathbf{w}_r, \gamma_{k}, \omega_{k})$ is given in (\ref{FP_2}).
Note that problem (P11) is still difficult to solve due to the non-convex objective (\ref{P11_obj}) and the coupling variables.

Therefore,
we again apply the BCA method to solve problem (P11).
Concretely,
the optimization variables are split into three blocks,
i.e.,
$\{\mathbf{w}_c, \mathbf{w}_r\}$,
$\{\gamma_{k}\}$,
and
$ \{\omega_{k}\}$,
and the objective in (P11) is maximized by blockwise alternating optimization: 
at each iteration one block is optimized while the remaining blocks are fixed, 
and this cycle is repeated until convergence.

\subsection{Optimizing Auxiliary Variables}

With other variables being fixed, 
the auxiliary variables 
$\gamma_{k}$
and
$ \omega_{k}$ are updated analytically by 
(\ref{FP_1_solution}) and (\ref{FP_2_solution}), respectively.

\subsection{Optimizing the Beamformers}

In this subsection,
we present the method to update the beamformers $\{\mathbf{w}_c, \mathbf{w}_r\}$.
When other variables are fixed,
the optimization problem of $\{\mathbf{w}_c, \mathbf{w}_r\}$
is meant to solve the following problem
\begin{subequations}
\begin{align}
\textrm{(P12)}:&\mathop{\textrm{max}}
\limits_{\mathbf{w}_c, \mathbf{w}_r}\
\mathbf{w}_c^H \mathbf{B}_{5}\mathbf{w}_c + \mathbf{w}_r^H\mathbf{B}_{6}\mathbf{w}_r
\label{P12_obj}\\
\textrm{s.t.}\ 
&  - \mathbf{w}_c^H \mathbf{B}_{1,k}\mathbf{w}_c - \mathbf{w}_r^H\mathbf{B}_{2,k}\mathbf{w}_r\label{P12_c_0}\\
&+
2\text{Re}\{ \mathbf{b}_{1,k}^H\mathbf{w}_c \} + c_{1,k}\geq R_{th}, \forall k \in \mathcal{K}, \nonumber\\
& \mathbf{w}_{c}^H \bar{\mathbf{A}}_{c,n} \mathbf{w}_{c} 
+ \mathbf{w}_{r}^H \bar{\mathbf{A}}_{r,n}\mathbf{w}_{r} \leq P_t, \forall n \in \mathcal{N}, \label{P12_c_1}
\end{align}
\end{subequations}
\noindent
where the objective function in (\ref{P12_obj}) and the rate constraint in (\ref{P12_c_0}) 
are obtained from (\ref{beampattern_trans}) and (\ref{Obj_trans}), respectively.
The above problem (P12) is still difficult to solve due to the non-convex objective function (\ref{P12_obj}).

Therefore,
following the linearization arguments in Sec. III-C,
the non-convex terms 
$\mathbf{w}_c^H \mathbf{B}_{5}\mathbf{w}_c$
and
$\mathbf{w}_r^H\mathbf{B}_{6}\mathbf{w}_r$
 can be replaced by  (\ref{SOCP_MM_1}) and (\ref{SOCP_MM_2}), respectively.
Thus,
problem (P12) can be reexpressed as
\begin{subequations}
\begin{align}
\textrm{(P13)}:&\mathop{\textrm{max}}
\limits_{\mathbf{w}_c, \mathbf{w}_r}\
\mathbf{w}_{c,0}^H \mathbf{B}_{5}\mathbf{w}_{c,0}
\!+\!
2\text{Re}\{ \mathbf{w}_{c,0}^H \mathbf{B}_{5}(\mathbf{w}_{c}\!\! -\!\! \mathbf{w}_{c,0}  )  \} \label{P13_obj}\\
& + \mathbf{w}_{r,0}^H \mathbf{B}_{6}\mathbf{w}_{r,0}
+
2\text{Re}\{ \mathbf{w}_{r,0}^H \mathbf{B}_{6}(\mathbf{w}_{r}\! -\! \mathbf{w}_{r,0}  )  \} \nonumber
\\
\textrm{s.t.}\ 
&  - \mathbf{w}_c^H \mathbf{B}_{1,k}\mathbf{w}_c - \mathbf{w}_r^H\mathbf{B}_{2,k}\mathbf{w}_r\label{P13_c_0}\\
&+
2\text{Re}\{ \mathbf{b}_{1,k}^H\mathbf{w}_c \} + c_{1,k}\geq R_{th}, \forall k \in \mathcal{K}, \nonumber\\
& \mathbf{w}_{c}^H \bar{\mathbf{A}}_{c,n} \mathbf{w}_{c} 
+ \mathbf{w}_{r}^H \bar{\mathbf{A}}_{r,n}\mathbf{w}_{r} \leq P_t, \forall n \in \mathcal{N}. \label{P13_c_1}
\end{align}
\end{subequations}

Problem (P13) is convex and can be solved by CVX.
The algorithm to solve the beampattern gain maximization problem, i.e., (P1), is specified in Alg. \ref{alg:3}.

\begin{algorithm}[t]
\caption{Solving the Problem (P1)}
\label{alg:3}
\begin{algorithmic}[1]
\STATE {initialize}
$\mathbf{w}_c^{(0)}$,
$\mathbf{w}_r^{(0)}$,
and
$t=0$;
\REPEAT
\STATE update $\{\gamma_k^{(t+1)}\}$ and $\{\omega_k^{(t+1)}\}$ by (\ref{FP_1_solution}) and (\ref{FP_2_solution}), respectively;
\STATE update $\mathbf{w}_c^{(t+1)}$ and  $\mathbf{w}_r^{(t+1)}$  by solving  (P13);
\STATE $t++$;
\UNTIL{$convergence$;}
\end{algorithmic}
\end{algorithm}

\subsection{Low-Complexity Solution}

Next,
we further propose a more efficient solution to solve problem (P13).
First,
we rewrite (P13) in a compact form as follows
\begin{subequations}
\begin{align}
\textrm{(P14)}:&\mathop{\textrm{min}}
\limits_{\mathbf{w}_{cr}}
- 2\text{Re}\{ \mathbf{b}_{4}^H\mathbf{w}_{cr}  \} + \bar{c}_{3} \label{P14_obj}
\\
\textrm{s.t.}\ 
&   \mathbf{w}_{cr}^H \bar{\mathbf{B}}_{7,k}\mathbf{w}_{cr} 
- 2\text{Re}\{ \bar{\mathbf{b}}_{3,k}^H\mathbf{w}_{cr} \} - \bar{c}_{1,k}\leq 0,   \label{P14_c_0}\\
& \mathbf{w}_{cr}^H \bar{\mathbf{A}}_{cr,n} \mathbf{w}_{cr}  \leq P_t, \forall n \in \mathcal{N}, \label{P14_c_1}
\end{align}
\end{subequations}
where
\begin{align}
&\bar{c}_{3} \triangleq ( \mathbf{w}_{cr,0}^H \mathbf{B}_{8} \mathbf{w}_{cr,0} )^{\ast},
\bar{\mathbf{B}}_{7,k} \triangleq \text{blkdiag}( \mathbf{B}_{1,k}, \mathbf{B}_{2,k}),\\
&\bar{\mathbf{b}}_{3,k} \triangleq [\mathbf{b}_{1,k}^T,\mathbf{0}^T  ]^T,
\bar{c}_{1,k} \triangleq {c}_{1,k} - R_{th},\nonumber
\end{align}
and other coefficients are given in Sec. III-D.

By introducing $K+1$ copies of the variable $\mathbf{w}_{cr}$,
i.e., $\mathbf{u}_{k} = \mathbf{w}_{cr}$, $\forall k \in \bar{\mathcal{K}} \triangleq  \{0\}\cup \mathcal{K}$,
we can decouple the $K+1$ constraints of (P14) as follows
\begin{subequations}
\begin{align}
\textrm{(P15)}:&\mathop{\textrm{min}}
\limits_{\mathbf{w}_{cr}, \{\mathbf{u}_{k}\}  }
- 2\text{Re}\{ \mathbf{b}_{4}^H\mathbf{w}_{cr}  \} + \bar{c}_{3} \label{P15_obj}
\\
\textrm{s.t.}\ 
&   \mathbf{u}_{k}^H \bar{\mathbf{B}}_{7,k}\mathbf{u}_{k}\!
-\! 2\text{Re}\{ \bar{\mathbf{b}}_{3,k}^H\mathbf{u}_{k} \}\! -\! \bar{c}_{1,k}\leq 0, \forall k \in \mathcal{K}, \label{P15_c_0}\\
& \mathbf{u}_{0}^H \bar{\mathbf{A}}_{cr,n} \mathbf{u}_{0}  \leq P_t, \forall n \in \mathcal{N}, \label{P15_c_1}\\
& \mathbf{w}_{cr} = \mathbf{u}_{k}, \forall k \in \bar{\mathcal{K}}. \label{P15_c_2}
\end{align}
\end{subequations}

{
In the following,
we again use the ADMM method \cite{ref_ADMM} to solve problem (P15).
First,
by relaxing the equality constraints (\ref{P15_c_2}) and penalizing them in the objective,
the AL function of problem (P15) can be represented as
\begin{align}
&\mathcal{L}( \mathbf{w}_{cr}, \{\mathbf{u}_k\}, \mathbf{u}_0, \{\boldsymbol{\lambda}_k\} )
=
- 2\text{Re}\{ \mathbf{b}_{4}^H\mathbf{w}_{cr}  \} + \bar{c}_{3}\\
&+ \frac{\rho_1}{2} {\sum}_{k=0}^{K} \Vert\mathbf{w}_{cr} - \mathbf{u}_{k}\Vert_2^2
+ {\sum}_{k=0}^{K} \text{Re}\{ \boldsymbol{\lambda}_k^H( \mathbf{w}_{cr} - \mathbf{u}_{k} ) \} ,\nonumber
\end{align}
where $\{\boldsymbol{\lambda}_k\}$ are the Lagrangian multipliers and  $\rho_1$ is a positive constant.
Then,
problem (P15) can be rewritten as
\begin{subequations}
\begin{align}
\textrm{(P16)}:&\mathop{\textrm{min}}
\limits_{\mathbf{w}_{cr}, \{\mathbf{u}_k\}, \mathbf{u}_0, \{\boldsymbol{\lambda}_k\} } \
\mathcal{L}( \mathbf{w}_{cr}, \{\mathbf{u}_k\}, \mathbf{u}_0, \{\boldsymbol{\lambda}_k\} ) \label{P16_obj}
\\
\textrm{s.t.}\ 
&   \mathbf{u}_{k}^H \bar{\mathbf{B}}_{7,k}\mathbf{u}_{k}\!
-\! 2\text{Re}\{ \bar{\mathbf{b}}_{3,k}^H\mathbf{u}_{k} \}\! -\! \bar{c}_{1,k}\leq 0, \forall k \in \mathcal{K}, \label{P16_c_0}\\
& \mathbf{u}_{0}^H \bar{\mathbf{A}}_{cr,n} \mathbf{u}_{0}  \leq P_t, \forall n \in \mathcal{N}. \label{P16_c_1}
\end{align}
\end{subequations}}

Next,
following the ADMM framework \cite{ref_ADMM} and using the BCD method,
we update variables $\mathbf{w}_{cr}$, $\{\mathbf{u}_k\}$, $\mathbf{u}_0$, and $ \{\boldsymbol{\lambda}_k\} $ sequentially.

\uwave{\textit{1) Update $\mathbf{w}_{cr}$:}}
When other variables are given,
the subproblem w.r.t. the variable $\mathbf{w}_{cr}$ is given as
\begin{subequations}
\begin{align}
\textrm{(P17)}:&\mathop{\textrm{min}}
\limits_{\mathbf{w}_{cr}} \
\mathbf{w}_{cr}^H  \mathbf{B}_9 \mathbf{w}_{cr} - 2\text{Re}\{\mathbf{b}_6^H \mathbf{w}_{cr} \} + c_3
\label{P17_obj}
\end{align}
\end{subequations}
where
\begin{align}
&\mathbf{B}_9 \triangleq \frac{\rho_1}{2}(K+1)\mathbf{I},\\
&\mathbf{b}_6 \triangleq \mathbf{b}_4 - {\sum}_{k=0}^{K} \frac{1}{2} \boldsymbol{ \lambda }_k 
+ {\sum}_{k=0}^{K} \frac{\rho_1}{2} \mathbf{u}_k , \nonumber\\
&c_3 \triangleq \bar{c}_3  + \frac{\rho_1}{2}( {\sum}_{k=0}^{K} \Vert \mathbf{u}_k \Vert_2^2)
- {\sum}_{k=0}^{K} \text{Re}\{ \boldsymbol{ \lambda }_k^H \mathbf{u}_k \}.\nonumber
\end{align}

{Problem (P17) is an unconstrained convex quadratic problem,
and its closed-form solution can be readily given as}
\begin{align}
\mathbf{w}_{cr}^{\star} = \mathbf{B}_9^{-1}\mathbf{b}_6. \label{Beam_w_closed_solution}
\end{align}

\uwave{\textit{2) Update $\{\mathbf{u}_{k}\}$:}}
The update of the variables $\{\mathbf{u}_{k}\}$ is reduced to solving the following problem
\begin{subequations}
\begin{align}
\textrm{(P18)}:&\mathop{\textrm{min}}
\limits_{ \{\mathbf{u}_k\} } \
{\sum}_{k=1}^{K} (\mathbf{u}_k^H\mathbf{u}_k -  2\text{Re}\{ \mathbf{b}_{7,k}^H \mathbf{u}_k \} ) + c_4
\label{P18_obj}
\\
\textrm{s.t.}\ 
&   \mathbf{u}_{k}^H \bar{\mathbf{B}}_{7,k}\mathbf{u}_{k}\!
-\! 2\text{Re}\{ \bar{\mathbf{b}}_{3,k}^H\mathbf{u}_{k} \}\! -\! \bar{c}_{1,k}\leq 0, \forall k \in \mathcal{K}, \label{P18_c_0}
\end{align}
\end{subequations}
where $c_4$ is a constant term and $\mathbf{b}_{7,k} \triangleq  ( \mathbf{w}_{cr} + \frac{1}{\rho_1}\boldsymbol{\lambda}_k ) $.

Obviously,
(P18) can be decomposed into $K$ independent subproblems,
and the $k$-th subproblem is given as
\begin{subequations}
\begin{align}
\textrm{(P18$_k$)}:&\mathop{\textrm{min}}
\limits_{ \mathbf{u}_k } \
\mathbf{u}_k^H\mathbf{u}_k - 2\text{Re}\{ \mathbf{b}_{7,k}^H \mathbf{u}_k \} 
\label{P18_1_obj}
\\
\textrm{s.t.}\ 
&   \mathbf{u}_{k}^H \bar{\mathbf{B}}_{7,k}\mathbf{u}_{k}\!
-\! 2\text{Re}\{ \bar{\mathbf{b}}_{3,k}^H\mathbf{u}_{k} \}\! -\! \bar{c}_{1,k}\leq 0. \label{P18_1_c_0}
\end{align}
\end{subequations}
Problem (P18$_k$) can be efficiently solved by leveraging Lemma \ref{lemma_1} in Sec. III-D.

\uwave{\textit{3) Update $\mathbf{u}_{0}$:}}
We proceed to study the optimization of $\mathbf{u}_{0}$, 
and the corresponding problem is given as
\begin{subequations}
\begin{align}
\textrm{(P19)}:&\mathop{\textrm{min}}
\limits_{ \mathbf{u}_0 } \
\mathbf{u}_{0}^H \mathbf{u}_{0} - 2\text{Re}\{ \mathbf{b}_{8}^H\mathbf{u}_{0} \} + c_5
\label{P19_obj}\\
\textrm{s.t.}\ 
&\mathbf{u}_{0}^H \bar{\mathbf{A}}_{cr,n} \mathbf{u}_{0}  \leq P_t, \forall n \in \mathcal{N}, \label{P19_c_0}
\end{align}
\end{subequations}
where $c_5$ is a constant term and $\mathbf{b}_{8} \triangleq  ( \mathbf{w}_{cr} + \frac{1}{\rho_1}\boldsymbol{\lambda}_0 ) $.

Note that the problem (P19)  can be split into $N$ subproblems.
The subproblem is formulated as 
\begin{subequations}
\begin{align}
\textrm{(P20)}:&\mathop{\textrm{min}}
\limits_{ \mathbf{u}_{0,n} } \
\mathbf{u}_{0,n}^H \mathbf{u}_{0,n} - 2\text{Re}\{ \mathbf{b}_{8,n}^H\mathbf{u}_{0,n} \}
\label{P20_obj}\\
\textrm{s.t.}\ 
&\mathbf{u}_{0,n}^H \mathbf{u}_{0,n}  \leq P_t, \label{P20_c_0}
\end{align}
\end{subequations}
where
\begin{align}
\mathbf{u}_{0,n} \triangleq &[\mathbf{u}_{0}(n),\mathbf{u}_{0}(n+N), \cdots, \mathbf{u}_{0}(n+ (K-1) N), \\
& \cdots, \mathbf{u}_{0}(n+ (K+N-1) N) ]^T, \nonumber\\  
\mathbf{b}_{8,n} \triangleq &[\mathbf{b}_{8}(n),\mathbf{b}_{8}(n+N), \cdots, \mathbf{b}_{8}(n+ (K-1) N), \nonumber\\
& \cdots, \mathbf{b}_{8}(n+ (K+N-1) N) ]^T.\nonumber
\end{align}

{We observe that problem (P20) has the same formulation as problem (P10). 
Problem (P20) can be efficiently solved by the Lagrange multiplier method in a similar manner. 
To avoid repetition, 
the detailed derivation is omitted here, 
and the update of $\mathbf{u}_{0,n}$ can be obtained 
by following the same procedure as that for problem (P10), i.e., (\ref{P10_1})$-$(\ref{P10_2}).}

\uwave{\textit{4) Update $\{\boldsymbol{\lambda}_k\}$:}}
{Finally,
the dual variables $\{\boldsymbol{\lambda}_k\}$ can be independently updated via the gradient ascent method,
which can be formulated as
\begin{align}
\boldsymbol{\lambda}_k^{(t+1)} := \boldsymbol{\lambda}_k^{(t)} + \rho_1(\mathbf{w}_{cr}^{(t+1)}-\mathbf{u}_k^{(t+1)}),
\forall k \in \bar{\mathcal{K}}.\label{ADMM_rho_1}
\end{align}}

The above ADMM-based algorithm to solve problem (P15) is summarized in Alg. \ref{alg:4}.

\begin{algorithm}[t]
\caption{The ADMM-Based Method to Solve (P15)}
\label{alg:4}
\begin{algorithmic}[1]
\STATE {initialize}
$\mathbf{w}_{cr}^{(0)}$,
$\{\mathbf{u}_k^{(0)}\}$,
$\{\boldsymbol{\lambda}_k^{(0)}\}$,
and
$t=0$;
\REPEAT
\STATE update $\mathbf{w}_{cr}^{(t+1)}$ by equation (\ref{Beam_w_closed_solution});
\FOR{ $k = 1:K$ }
\STATE update $\mathbf{u}_{k}^{(t+1)}$ by solving (P18$_k$);
\ENDFOR
\FOR{ $n = 1:N$ }
\STATE update $\mathbf{u}_{0,n}^{(t+1)}$ by solving (P20);
\ENDFOR
\STATE update $\boldsymbol{\lambda}^{(t+1)}$ by equations (\ref{ADMM_rho_1});
\STATE $t++$;
\UNTIL{$convergence$;}
\end{algorithmic}
\end{algorithm}

\subsection{Convergence and Complexity Analysis}

\textit{1) Convergence Analysis:}
{
Since Alg. \ref{alg:3} is designed within the MM framework \cite{ref_MM}, \cite{ref_convergence},  
the original objective value of problem (P1) is monotonically non-decreasing over the iterations. 
According to \cite{ref_convergence},  
Alg. 3 is guaranteed to converge to a stationary point of the original problem (P1).}

\textit{2) Complexity Analysis:}
{Since Alg. \ref{alg:3} and Alg. \ref{alg:4} share similar algorithm structures with Alg. \ref{alg:1} and Alg. \ref{alg:2}, 
respectively,
their complexity analysis follows analogously to that of the sum-rate maximization problem.
Therefore, the detailed derivation is omitted to avoid repetition.}

\section{Numerical Results}

\begin{figure}[t]
	\centering
	\includegraphics[width=.30\textwidth]{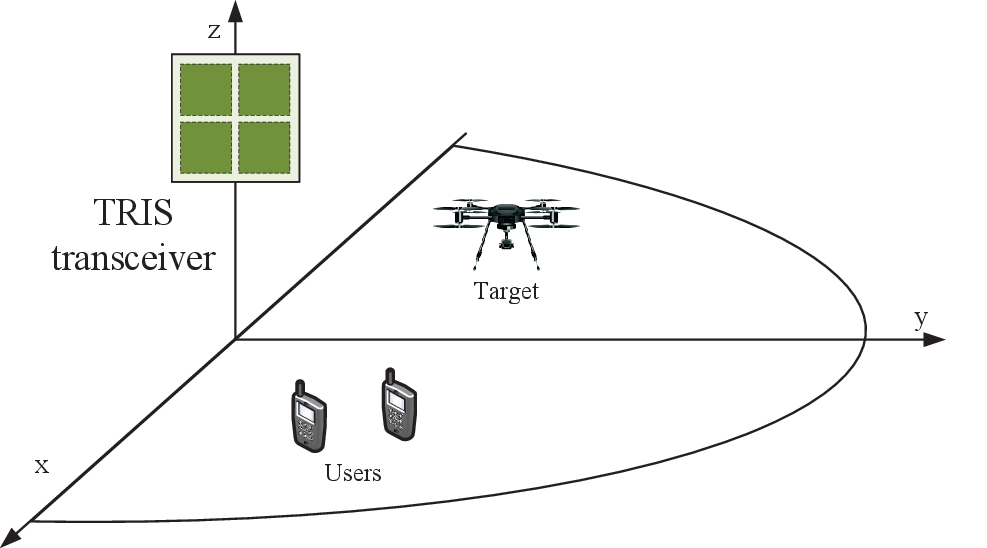}
	\caption{The experiment scenario model.}
	\label{fig.2}
\end{figure}

In this section, 
we provide extensive numerical results to verify the system performance in the TRIS transceiver-enabled ISAC system and the effectiveness of the proposed algorithms.
As shown in Fig. \ref{alg:2},
we consider an ISAC system consisting of a TRIS transceiver with $N=16$ units, $K=2$ users, and one point-like sensing target.
In our experiment,
the TRIS transceiver is located at $(0,0,4.5)$ m in three-dimensional (3D) space. 
The mobile users are randomly placed in a sector spanning $30$~m-$50$~m from the origin and are set at an elevation of $1.5$ m.
The element spacing is set to half the wavelength of the carrier.
The path-loss exponents of the TRIS transceiver-user and the TRIS transceiver-sensing target links are set as $3.2$ and $2.2$, respectively.
The maximum power of each TRIS unit is $10$ dBm.
The noise power at all receivers is the same, i.e., $\sigma_{c,k}^2 = -90$ dBm, $\forall k \in \mathcal{K}$ \cite{ref_noise}.

\subsection{Sum-Rate Maximization}
In this subsection, 
we present simulation results for the sum-rate maximization problem.

\begin{figure}[t]
	\centering
	\includegraphics[width=.43\textwidth]{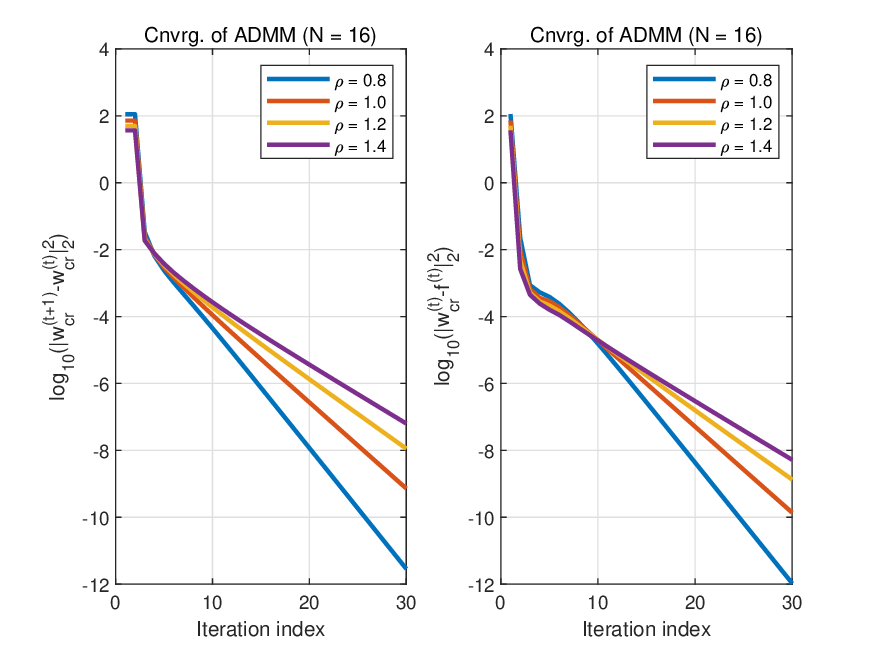}
	\caption{Convergence of ADMM algorithm.}
	\label{fig.3}
\end{figure}

\begin{figure}[t]
	\centering
	\includegraphics[width=.43\textwidth]{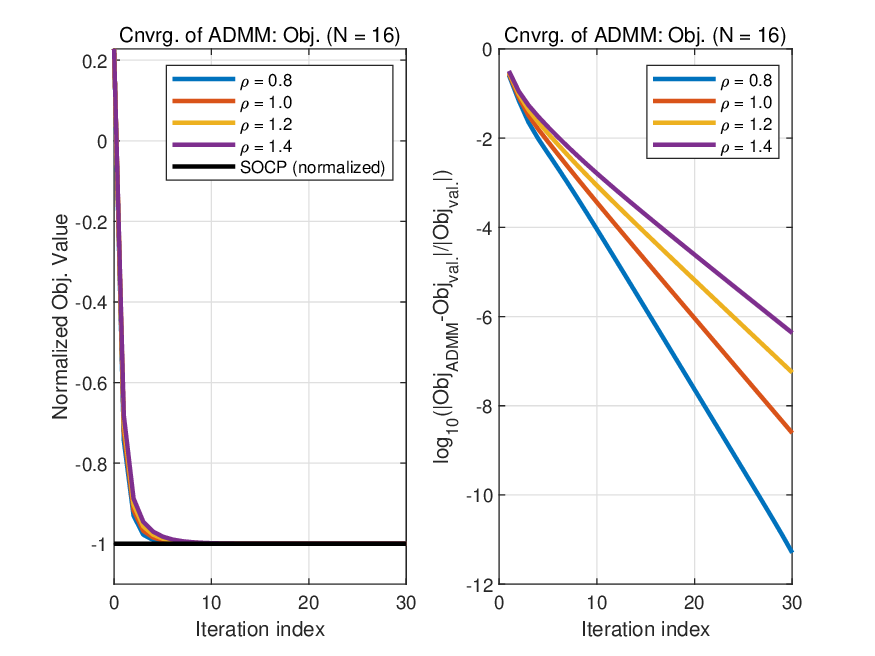}
	\caption{Convergence of objective value of ADMM method.}
	\label{fig.4}
\end{figure}

First,
Fig. \ref{fig.3} and Fig. \ref{fig.4} jointly demonstrate the convergence behaviors of Alg. \ref{alg:2},
which updates beamforming using the ADMM method.
Under different values of the parameter $\rho$,
the variation in $\mathbf{w}_{cr}$ itself and the difference between $\mathbf{w}_{cr}$ and $\mathbf{f}$ across loop iterations are plotted in the left and right parts of Fig. \ref{alg:3}, respectively.
As illustrated in the figure,
the ADMM-based algorithm generally reaches convergence in about 30 iterations 
when the condition is well satisfied (i.e., the residual drops below $10^{-6}$).
Moreover,
by comparing these curves for different values of $\rho$,
the figure also shows how the parameter $\rho$ influences the convergence rate, the residual magnitude, and the numerical stability of the ADMM-based algorithm.

Fig. \ref{fig.4} illustrates the objective convergence behavior produced by Alg. \ref{alg:2}.
To guarantee fairness, 
both the SOCP-based and the low-complexity methods begin from an identical initial point in every channel realization.
In the left subfigure, 
the evolution of the objective value is plotted for different values of the parameter $\rho$.
The black line indicates the objective value of the problem (P5) computed using CVX.
The right subfigure shows, 
in the log domain, 
the difference between the objective values produced by CVX and Alg. \ref{alg:2}, 
which makes it easy to observe how the residual evolves over iterations.
Generally,
as shown in Fig. \ref{fig.4},
Alg. \ref{alg:2} achieves a sufficiently accurate objective value rapidly.

\begin{table}[t]
\begin{small}
\centering
\caption{MATLAB Run Time to Solve (P5) (in Sec.)}
\begin{tabular}{|c|c|c|c|c|c|} \hline \label{Table_SOCP_vs_ADMM_1}
Alg. & $N$ = 4  & $N$ = 8 & $N$ = 12& $N$ = 16 & $N$ = 20      \\ \hline
SOCP & 0.2953   & 0.3466  & 0.4187  & 0.5683   & 0.8006      \\ \hline
ADMM & 0.0040   & 0.0102  &	0.0492  & 0.1378   & 0.3094      \\ \hline
\end{tabular}

\end{small}
\end{table}

In Table \ref{Table_SOCP_vs_ADMM_1}, 
we test the computational complexity of the analytical solution (i.e., Alg. \ref{alg:2}) and the SOCP-based algorithm.
The MATLAB execution time for CVX and Alg. \ref{alg:2} for different numbers of TRIS transceiver elements $N$ is shown in Table \ref{Table_SOCP_vs_ADMM_1}.
As shown in this table, 
the run time for both implementations increases as the number of TRIS transceiver units $N$ increases. 
Moreover, 
the run time of the analytical solution is roughly one to two orders of magnitude lower than that of the SOCP-based algorithm.

\begin{figure}[t]
	\centering
	\includegraphics[width=.43\textwidth]{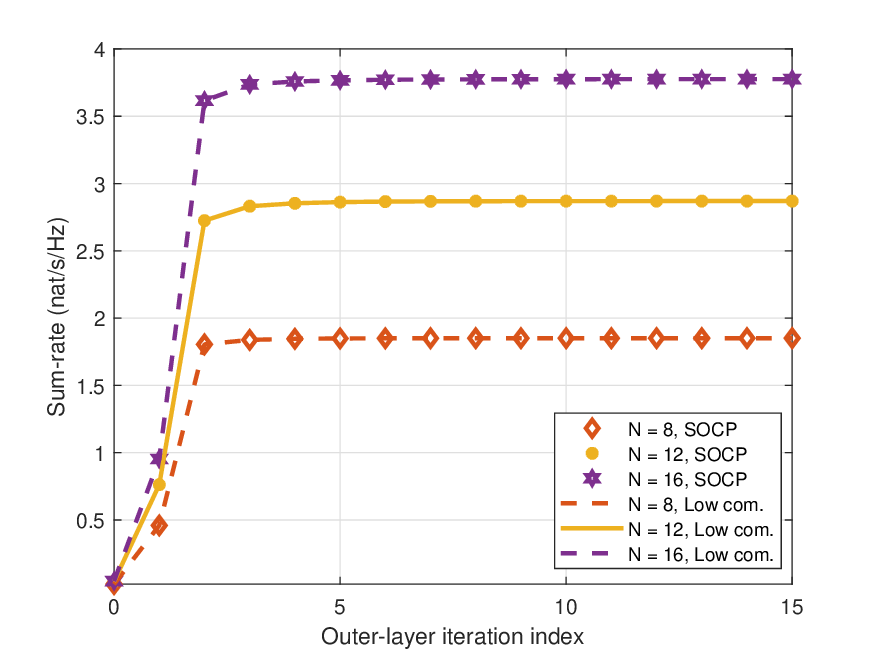}
	\caption{Convergence of algorithms.}
	\label{fig.6}
\end{figure}

The overall convergence behaviors of the proposed algorithms for solving the sum-rate maximization problem (P0) are shown in Fig. \ref{fig.6},
where the SOCP-based solution is compared with the low-complexity (labeled ``Low com.'') solution across different TRIS transceiver unit numbers.
To ensure a fair comparison, 
both methods begin from an identical initial point in each channel.
The figure demonstrates that, 
for any element number, 
the SOCP-based and analytic solutions produce almost the same objective values, 
confirming that the analytical approximation maintains the SOCP-based solution quality.
Furthermore, both algorithms converge quickly, with the sum-rate stabilizing within approximately 5 outer iterations.
With substantially reduced per-iteration run time (see Table \ref{Table_SOCP_vs_ADMM_1}) and comparable convergence performance, 
the low-complexity solution is more efficient overall.

\begin{figure}[t]
	\centering
	\includegraphics[width=.43\textwidth]{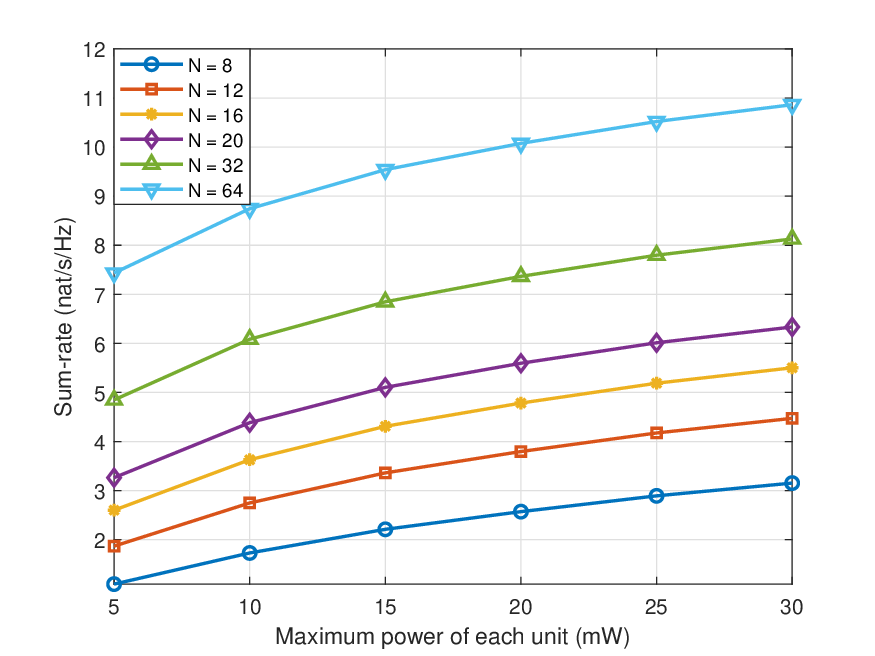}
	\caption{Sum-rate versus the maximum power of each TRIS unit.}
	\label{fig.7}
\end{figure}

Fig.~\ref{fig.7} shows the effect of the maximum transmit power per TRIS unit on the achieved sum-rate. 
In this experiment, the maximum transmit power per TRIS unit varies from 5 mW to 30 mW. 
As observed in Fig.~\ref{fig.7}, the sum-rate increases monotonically with the transmit power for all considered numbers of TRIS units. 
This behavior suggests that a higher per-unit power budget relaxes the per-unit power constraints and enlarges the feasible beamforming region. 
Consequently, the TRIS transceiver can allocate more energy to the communication users while still satisfying the sensing requirement, 
thereby improving the received SINR and the achievable sum-rate. 
In addition, larger TRIS arrays consistently achieve a higher sum-rate at each power level. 
This result further demonstrates that larger arrays offer stronger array gain and greater beamforming flexibility. 
Although the sum-rate continues to improve with transmit power, 
the slope of the curves gradually decreases in the high-power regime.

\subsection{Beampattern Gain Maximization}
In this subsection,  we provide simulation results for the beampattern gain maximization problem.

\begin{figure}[t]
	\centering
	\includegraphics[width=.43\textwidth]{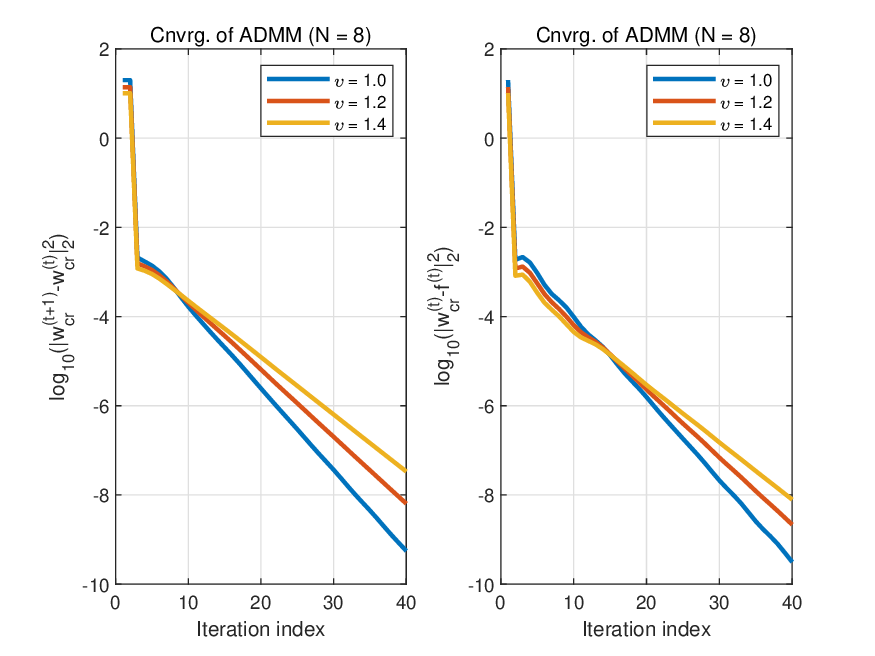}
	\caption{Convergence of ADMM algorithm.}
	\label{fig.9}
\end{figure}

\begin{figure}[t]
	\centering
	\includegraphics[width=.43\textwidth]{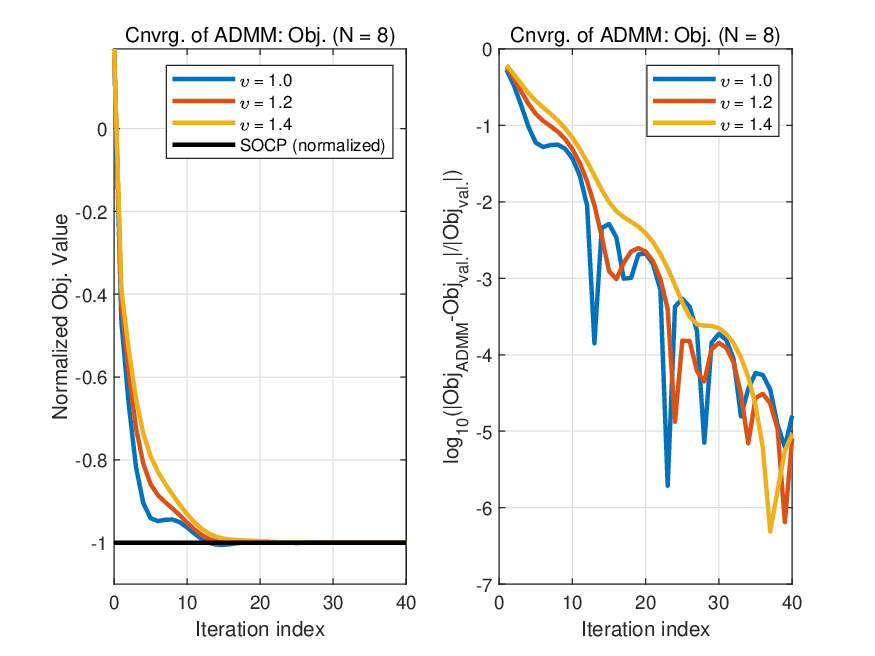}
	\caption{Convergence of objective value of ADMM method.}
	\label{fig.10}
\end{figure}

First,
Fig. \ref{fig.9} and Fig. \ref{fig.10} jointly present the convergence behaviors of the proposed ADMM-based method for solving (P15).
We can observe that selecting a suitable penalty coefficient $\rho_1$ generally leads to a satisfactory rate of convergence.
This method usually converges within 30 to 40 iterations.
The equality constraints $\mathbf{w}_{cr}=\mathbf{u}_k$ are effectively enforced by 30 iterations, 
and the objective value generally reaches a precision on the order of $10^{-3}$ within 30 iterations.

\begin{table}[t]
\begin{small}
\centering
\caption{MATLAB Run Time to Solve (P15) (in Sec.)}
\begin{tabular}{|c|c|c|c|c|c|} \hline \label{Table_SOCP_vs_ADMM_2}
Alg. & $N$ = 4  & $N$ = 6 & $N$ = 8 & $N$ = 10 & $N$ = 12    \\ \hline
SOCP & 0.6530   & 0.6808  & 0.7612  & 0.8719   & 0.9945      \\ \hline
ADMM & 0.0317   & 0.0393  &	0.0496  & 0.0717   & 0.1224      \\ \hline
\end{tabular}

\end{small}
\end{table}

Besides, 
Table \ref{Table_SOCP_vs_ADMM_2} demonstrates a comparison of the run time of the SOCP and ADMM methods when solving (P15).
The MATLAB run time for varying numbers of TRIS elements (i.e., $N$) is summarized in Table \ref{Table_SOCP_vs_ADMM_2}.
Combining the results in Fig. \ref{fig.9}, Fig. \ref{fig.10}, and Table \ref{Table_SOCP_vs_ADMM_2}, 
we can conclude that the ADMM-based method is more efficient than the SOCP-based method.

\begin{figure}[t]
	\centering
	\includegraphics[width=.43\textwidth]{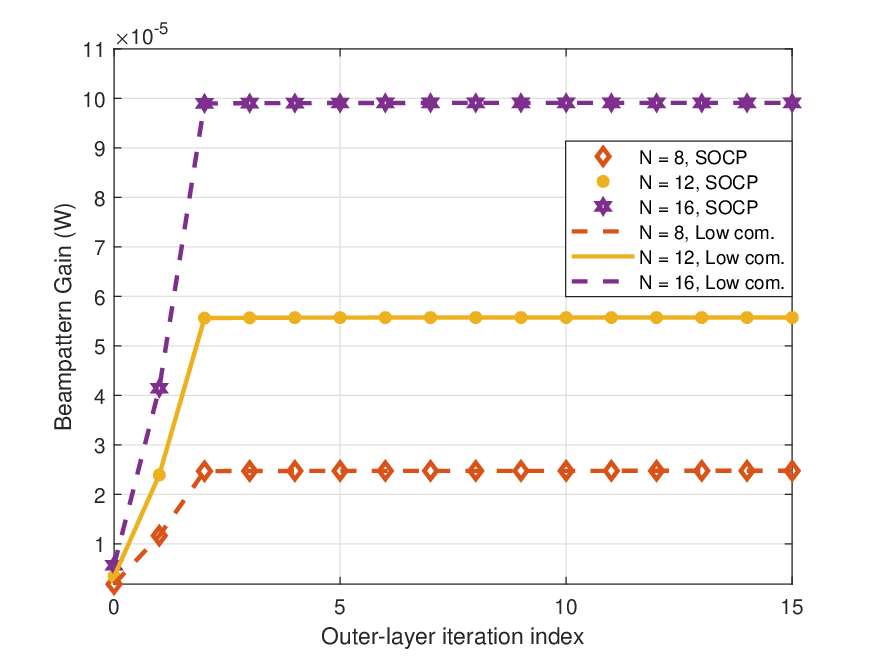}
	\caption{Convergence of algorithms.}
	\label{fig.11}
\end{figure}

Fig. \ref{fig.11} illustrates the convergence behaviors of the developed methods (i.e., the SOCP-based and Low
com. solutions) when solving the beampattern gain maximization problem (P1) under different numbers of TRIS units. 
From Fig. \ref{fig.11}, we can observe that both algorithms exhibit identical performance and generally achieve convergence within 5 iterations.
Moreover, it is also shown that a larger number of TRIS elements further improves the beampattern gain.

\begin{figure}[t]
	\centering
	\includegraphics[width=.43\textwidth]{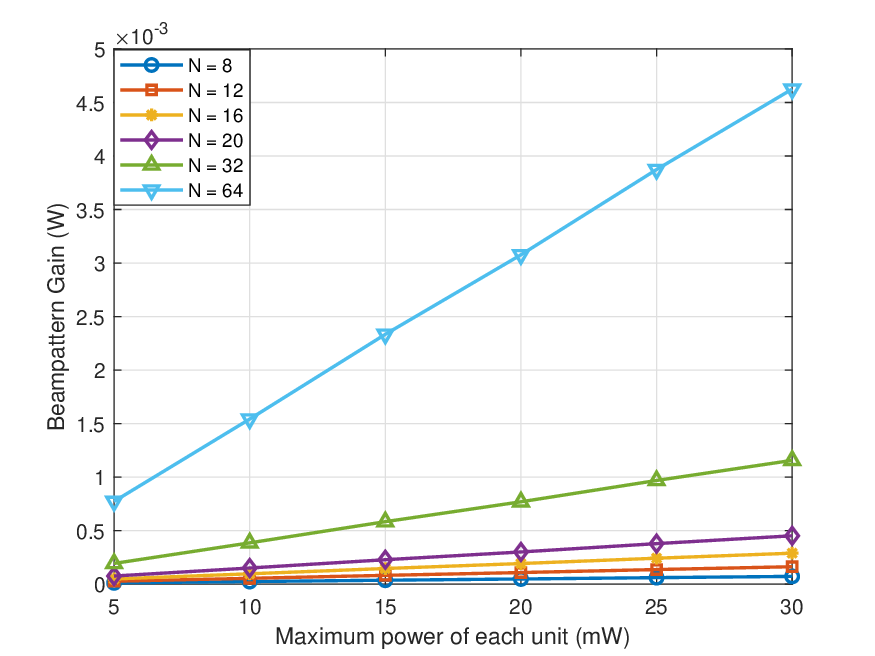}
	\caption{Beampattern gain versus the maximum power of each TRIS unit.}
	\label{fig.12}
\end{figure}

Finally, Fig. \ref{fig.12} illustrates the effect of the maximum transmit power of each TRIS unit on the beampattern gain. It can be observed that the beampattern gain increases steadily with the per-unit power budget for all considered TRIS array sizes. 
This result indicates that a higher available transmit power enables more energy to be concentrated toward the target direction, 
which improves the target illumination capability. 
In addition, larger TRIS arrays achieve a higher beampattern gain over the entire power range, 
reflecting their superior coherent combining capability and finer spatial control of the transmitted signal.

\section{Conclusions}

In this paper,
we propose the use of a TRIS transceiver to enable simultaneous communication and sensing in an ISAC system with multiple mobile users and one point-like target.
Specifically,
the communication and radar beamformers are jointly optimized to maximize the sum-rate/beampattern gain, 
subject to
the minimum beampattern gain/per-user rate requirement and per-unit power constraints of the TRIS transceiver .
Based on the FP and MM methods,
we successfully develop efficient solutions to solve the sum-rate/beampattern gain maximization problems.
In particular,
by leveraging the ADMM framework,
we propose two iterative closed-form algorithms for the sum-rate and beampattern gain maximization problems, respectively.
Numerical simulation results demonstrate that the proposed optimization algorithms can substantially enhance sum-rate and
beampattern gain, respectively.

\normalem


\end{document}